
%
%
%
%
%
%
%
%
%
\input amstex
\documentstyle{conm-p}

\def\hf{\frac{1}{2}}
\def\shf{{\tsize{\frac{1}{2}}}}
\def\sth{{\tsize{\frac{3}{2}}}}

\def\sfr{{\tsize{\frac{1}{4}}}}
\def\sot{{\tsize{\frac{1}{3}}}}
\def\stt{{\tsize{\frac{2}{3}}}}
\def\stf{{\tsize{\frac{3}{4}}}}
\def\sft{{\tsize{\frac{4}{3}}}}
\def\fif{{\tsize{\frac{5}{4}}}}
\def\sft{{\tsize{\frac{4}{3}}}}
\def\sei{{\tsize{\frac{1}{8}}}}
\def\ste{{\tsize{\frac{3}{8}}}}
\def\six{{\tsize{\frac{1}{16}}}}
\def\svx{{\tsize{\frac{7}{16}}}}
\def\sr2{{\tsize{\frac{1}{\sqrt 2}}}}
\def\hp{\hat p}

\def\bDel{{\bar\Delta}}
\def\bV{\bold V}
\def\cL{{\Cal L}}
\def\cR{{\Cal R}}

\def\tPhi{{\tilde{\Phi}}}
\def\tPsi{{\tilde{\Psi}}}
\def\bcV{{\bold{\Cal V}}}
\def\bCl{\bold{Cliff}}
\def\bCM{\bold{CM}}
\def\bVir{\bold{Vir}}
\def\BZ'{{\Bbb Z}+\shf}
\def\bvac'{\bold{vac'}}
\def\bvac{\bold{vac}}

\def\bi{\bold i}
\def\ds{\displaystyle}
\def\bk{\blacksquare}
\def\lra{\leftrightarrow}

\def\sk1{\vskip 10pt}

\def\smK{\sum_{0 \leq k \in {\Bbb Z}}}
\def\smQ{\sum_{0 \leq q \in {\Bbb Z}}}
\let\pr\proclaim
\let\epr\endproclaim

\topmatter
\title
Spinor Construction of the $c = 1/2$ Minimal Model
\endtitle

\rightheadtext{Spinor Construction of the $c = 1/2$ Minimal Model}

\author Alex J. Feingold, John F. X. Ries, Michael D. Weiner
\endauthor

\leftheadtext{FEINGOLD, RIES, WEINER}

\address
Dept. of Math. Sci.,
The State University of New York,
Binghamton, New York 13902-6000
\endaddress

\address
Math. Dept.,
East Carolina University,
Greenville, North Carolina 27858-4353
\endaddress

\address
Math. Dept.,
Broome Community College,
Binghamton, New York 13902-1017
\endaddress

\thanks
A.J.F. was partially supported by the National Security Agency under
Grant number MDA904-94-H-2019. The United States Government is authorized to
reproduce and distribute reprints notwithstanding any copyright notation
heron.
\endgraf
J.F.X.R. died July 6, 1993. This paper is dedicated to his memory.
\endgraf
This paper is in final form and no version of it will be submitted for
publication elsewhere.
\endthanks


\endtopmatter

\document

\noindent{\bf{1. Introduction}}

The representation theories of affine Kac-Moody Lie algebras and the Virasoro
algebra have been found to have interesting connections with many other
parts of mathematics and have played a vital role in the development of
conformal field theory in theoretical physics.(See \cite{BPZ}, \cite{GO},
\cite{BMP}, \cite{MS},
\cite{TK}, the Introduction and references in \cite{FLM}.) An
important contribution to Kac-Moody representation theory was the
rigorous mathematical construction of representations using vertex operators,
discovered independently by mathematicians and physicists, but known earlier
by physicists. Once the connection was discovered, an exciting dialog began
which has enriched both sides. The precise axiomatic definition of vertex
operator algebras (VOA's) by mathematicians \cite{FLM}
gave a rigorous foundation
to the algebraic aspects of conformal field theory called chiral algebras by
physicists. This theory includes, unifies and vastly extends the representation
theories of affine Kac-Moody algebras and the Virasoro algebra.

In the theory of VOA's one has the notion of a module and of intertwining
operators going between modules \cite{FHL}, \cite{F}. The main axiom for a VOA
is an identity called the Jacobi-Cauchy identity because it combines the usual
Jacobi identity for Lie algebras with the Cauchy residue formula for
rational functions whose possible poles are limited to three points, $0$, $1$
and $\infty$. A slight modification is needed for the appropriate definition
of a module, and intertwining operators are defined by a similar axiom
relating them to vertex operators. We believe that an important next step is
the understanding of a new kind of ``matrix'' Jacobi-Cauchy identity relating
any two intertwining operators. This would lead to a larger unifying structure,
incorporating a VOA, its modules and its intertwining operators.

There are several new features which appear when trying to understand
intertwining operators and the new kind of Jacobi-Cauchy identity they obey.
First one has to deal with fusion rules,
$${\Cal N}(M_1,M_2,M_3) = \dim({\Cal I}(M_1,M_2,M_3)),$$
which give the dimension of the
space of intertwining operators determined by a triple of modules. It is a
basic principle of VOA's that there is a one-to-one correspondence between
vectors $v$ in a simple VOA $V$ and vertex operators $Y_M(v,z)$ acting on an
irreducible $V$-module $M$. One can think of this as a map
$$Y_M(\cdot,z) : V\to End(M)[[z,z^{-1}]]$$
which obeys the various axioms defining a $V$-module. The fusion rule in this
case is always ${\Cal N}(V,M,M) = 1$ and one axiom normalizes $Y_M$ so it is
uniquely determined. Given three $V$-modules, $M_1$, $M_2$, $M_3$, one can
think of an intertwining operator as a map
$$Y(\cdot,z) : M_1\to Hom(M_2,M_3)\{\{z,z^{-1}\}\}$$
which obeys the axioms for intertwining operators. (The notation
$\{\{z,z^{-1}\}\}$ indicates rational powers of $z$.)
It is quite possible that the fusion rule is ${\Cal N}(M_1,M_2,M_3) = n > 1$,
and in that case one does not
have a one-to-one correspondence between vectors $w$ in module $M_1$ and
operators $Y(w,z)$ whose components send $M_2$ to $M_3$.
It would seem that this is a kind
of labeling problem, there not being enough ``copies'' of the vectors in
$M_1$ to distinguish the $n$ linearly independent intertwiners which could be
taken as a basis for the space of all intertwiners. It is also possible to
have four modules $M_1,\cdots,M_4$ with $M_3 \ne M_4$, with fusion rules
${\Cal N}(M_1,M_2,M_3) \geq 1$ and ${\Cal N}(M_1,M_2,M_4) \geq 1$. This also
indicates a labeling problem, showing the inadequacy of the notation
$Y(w_1,z)w_2$, where knowing that $w_i\in M_i$ still does not determine
which module contains the outcome.

Another new feature which appears is the nature of the correlation functions,
$$(Y(w_1,z_1)Y(w_2,z_2)w_3,w_4)\quad\hbox{and}\quad
(Y(Y(w_1,z)w_2,z_2)w_3,w_4),$$
made from two intertwiners. These are series which converge in certain
domains to functions which, after factoring out some rational powers of
$z_1$ and $z_2$, can be expressed as power series in $z_2/z_1$ and $z/z_2$
satisfying certain differential equations. One way of thinking of the usual
Jacobi-Cauchy identity for a VOA $V$ is as follows. The three series
$$(Y(v_1,z_1)Y(v_2,z_2)v_3,v_4),\qquad
(Y(v_2,z_2)Y(v_1,z_1)v_3,v_4)$$
and
$$(Y(Y(v_1,z_1-z_2)v_2,z_2)v_3,v_4)$$
converge in their respective domains,
$|z_1| > |z_2|$, $|z_2| > |z_1|$, and $|z_2| > |z_1-z_2|$, to the same rational
function $f(z_1,z_2)$ in the ring
$${\Cal R} = {\Bbb C}[z_1,z_1^{-1},z_2,z_2^{-1},(z_1 - z_2)^{-1}].$$
For fixed $z_2 \ne 0$, these are functions of $z_1$ with possible poles only at
$z_1=0$, $z_1=\infty$ and $z_1=z_2$. The Jacobi-Cauchy identity is just the
statement of the Cauchy residue theorem applied to
$f(z_1,z_2)g(z_1,z_2)$, where $g(z_1,z_2) \in {\Cal R}$ is arbitrary, and the
residues at each of the three points are computed from the three series
giving $f$ and appropriate series expansions of $g$.
But the kinds of correlation functions obtained from intertwiners
generally involve hypergeometric functions to which the treatment just
described does not apply.

The purpose of this paper is to show how both of these new features can be
handled in a simple but nontrivial example which may have important
applications. In the monograph \cite{FFR} we constructed certain vertex
operator
superalgebras (VOSA's) and their twisted modules from Clifford algebras and
their spinor representations. These ``fermionic'' constructions extend the
corresponding constructions of the orthogonal affine Kac-Moody algebras of
type $D_n^{(1)}$ in \cite{FF}. (See \cite{Fr1}, \cite{Fr2}.)
In \cite{W} vertex operator para-algebras were
constructed from the bosonic constructions of level $-\shf$ representations of
symplectic affine Kac-Moody algebras of type $C_n^{(1)}$. It is well
known to physicists, and follows immediately from the work in \cite{FFR}, that
a spinor construction from one fermion gives a VOSA and a twisted module for
it. These are known in the physics literature as the Neveu-Schwarz and Ramond
sectors, respectively, and they each decompose into two irreducible modules
for the Virasoro algebra with central charge $c=\shf$. Using the usual
notation for labeling Virasoro modules, the two components
into which the Neveu-Schwarz sector decomposes are labeled by $h=0$ and
$h=\shf$, and the two coming from the Ramond sector are both $h=\six$. It is
very interesting and important that in this construction one naturally has
two copies of the $h=\six$ Virasoro module. That enables us to define unique
intertwining operators for each vector in the Ramond sector, such that the
usual Ising fusion rules for just three modules are replaced by fusion rules
given by the group ${\Bbb Z}_4$. This behavior is just like the behavior of
vertex operator para-algebras defined in \cite{FFR}. (See also \cite{DL}.)
It means that the VOA $V$ and
its modules are indexed by a finite abelian group $\Gamma$ such that
$V = V_0$ ($0\in\Gamma$ is the identity element) and the fusion rules are
${\Cal N}(V_a,V_b,V_c) = 1$ if $a+b=c$ in $\Gamma$, zero otherwise.
The important question is whether this is a rare special situation, or if
there are other natural constructions of VOA's where multiple copies of the
modules allow unique labeling of intertwining operators and where fusion
rules are replaced by a group law.

In the example studied here we find that the hypergeometric functions to
which the correlation function converge, come in pairs, as bases for the two
dimensional spaces of solutions of certain differential equations. In order
to relate them to each other, we must use Kummer's quadratic transformation
formula. This involves certain substitutions which lift the correlation
functions of $t = z_2/z_1$ to functions of $x$ on a four-sheeted covering of
the $t$-sphere, branched at $t = 0,1,\infty$ with possible poles only at
$x = \alpha\in\{0,\infty, 1,-1,\bi,-\bi\}$.
We may then find the $2\times 2$ matrices
$B_\alpha$ which relate the $2\times 4$ matrices of transformed correlation
functions at $x = 0$ to those at $x = \alpha$. The matrix-valued
functions have possible poles only at those six points, so after multiplying
the matrix-valued functions by any function $g(x)$ in the ring
$${\Cal R}_x = {\Bbb C}[x,x^{-1}(x^4-1)^{-1}]$$
the Cauchy residue theorem gives that the sum of their residues at the six
points adds up to zero. Expressing the residues in terms of the series and
expanding the function $g(x)$ as an appropriate series, we get a ``matrix''
Jacobi-Cauchy identity. It is, of course, actually a generating function of
identities, equivalent to infinitely many identities for the components of
the intertwining operators. It may take some time to sort out which ones are
the most important, but we can already point out some interesting ones.

A very interesting aspect of this example is the structure of the matrices
$B_\alpha$. In order to find those matrices one must choose some linear
fractional transformations relating the variable $x = x_0$ to the variables
$x_\alpha$ which are local variables at the poles. There are several sign
choices which can be made, and these correspond to choices of braiding. It
is not so surprising that these matrices provide a representation of a braid
group. The details of this aspect of the example have not been completely
worked out yet, and will be studied later. There should be an interesting
connection with the work in \cite{MaS}.

As a final motivation for the detailed study of this special example, we
would like to mention the possible application to a spinor construction
of the moonshine module for the monster group. It has been noted by Dong,
Mason and Zhu \cite{DMZ} that there are 48 commuting $c=\shf$ Virasoro algebras
in the moonshine module $V^\natural$ considered as a VOA. It means that
$V^\natural$ decomposes into a sum of tensor products of 48 Virasoro modules,
each of which is one of the $h = 0$, $h = \shf$ or $h = \six$ modules.
Although some information is known about this decomposition, the complete
picture is not clear. But since there is a spinor construction of each of
these modules, there is some hope that a spinor construction of $V^\natural$
is possible. In fact, we hope that the new light we have shed on the $c =
\shf$ minimal model will be of help in achieving that goal.
\vskip 15pt

\noindent{\bf{2. Construction of Vertex Operator Superalgebra and Module}}
\sk1

Let $E = {\Bbb C}e$ with $\langle e,e\rangle = 2$ and let $Z = {\Bbb Z}$
or $Z = \BZ'$. Let $E(Z)$ be the vector space with basis
$\{e(m)\ |\ m \in Z\}$ and the symmetric form
$$\langle e(m),e(n)\rangle = \langle e,e\rangle\delta_{m,-n} =
2\delta_{m,-n}.$$
Let $\bCl(Z)$ be the Clifford algebra generated by $E(Z)$ and that form.
Let $E(Z) = E(Z)^+ \oplus E(Z)^-$ be the polarization where
$E(Z)^+$ is spanned by $\{e(m)\ |$ $\ 0 < m \in Z\}$ and
$E(Z)^-$ is spanned by $\{e(m)\ |\ 0 \geq m \in Z\}$. Define ${\goth I}(Z)$
to be the left ideal in $\bCl(Z)$ generated by $E(Z)^+$, so that
$$\bCM(Z) = \bCl(Z)/{\goth I}(Z)$$
is a left $\bCl(Z)$-module. One has the parity decomposition
$$\bCM(Z) = \bCM(Z)^0 \oplus \bCM(Z)^1$$
where $\bCM(Z)^i$, $i = 0,1$, is the subspace with basis
$$\{e(-m_1)\hdots e(-m_r)\bvac(Z)\ |\ m_1>\hdots > m_r\geq  0,\
m_1,\hdots,m_r\in Z,\ r\equiv i\hbox{ mod }2 \}$$
and $\bvac(Z) = 1 + {\goth I}(Z)$.
Define the vacuum space
$${\bold{VAC}}(Z) = \{v\in\bCM(Z)\ |\ E(Z)^+\cdot v = 0\}.$$
Then ${\bold{VAC}}(\BZ')$ is one-dimensional, spanned by
$$\bvac = \bvac(\BZ') = 1 + {\goth I}(\BZ'),$$
and ${\bold{VAC}}({\Bbb Z})$ is two-dimensional, spanned by
$$\bvac' = \bvac({\Bbb Z}) = 1 + {\goth I}({\Bbb Z}) \qquad\hbox{and}
\qquad e(0)\bvac'.$$
It is easy to see that $\bCM(\BZ')$ is an irreducible $\bCl(\BZ')$-module,
but $\bCM({\Bbb Z})$ decomposes into two irreducible
$\bCl({\Bbb Z})$-modules. Note that
$$\bvac'_+ = \bvac' + e(0)\bvac' \qquad \hbox{and}\qquad
\bvac'_- = \bvac' - e(0)\bvac'$$
are eigenvectors for $e(0)$ with eigenvalues $+1$ and $-1$, respectively,
because the relations in $\bCl({\Bbb Z})$ give $e(0)^2 = 1$. We then have
the alternative decomposition into two irreducible $\bCl({\Bbb Z})$-modules,
$$\bCM({\Bbb Z}) = \bCM({\Bbb Z})^+ \oplus \bCM({\Bbb Z})^-$$
where $\bCM({\Bbb Z})^\pm$ is the subspace with basis
$$\{e(-m_1)\hdots e(-m_r)\bvac'_\pm \ |\ m_1>\hdots > m_r > 0,\
m_1,\hdots,m_r\in {\Bbb Z} \}.$$

Later we will need the vector space isomorphism
$$\theta : \bCM({\Bbb Z}) \to  \bCM({\Bbb Z})$$
defined by
$$\theta(\bvac') = e(0)\bvac' \qquad \hbox{and}\qquad
\theta(e(m)v) = e(m)\theta(v)$$
for any $m\in {\Bbb Z}$ and $v\in\bCM({\Bbb Z})$. It is clear that $\theta$
is an involution switching $\bCM({\Bbb Z})^0$ with $\bCM({\Bbb Z})^1$, and
that $\bCM({\Bbb Z})^+$ and $\bCM({\Bbb Z})^-$ are the $+1$ and $-1$
eigenspaces, respectively, for $\theta$.

Define the $\bVir$  operators
$$L(k) = -\sfr \sum_{n\in Z} (n + \shf)\ {^\circ_\circ}
e(n)e(k - n) {^\circ_\circ} \ \hbox{ for }\ k \neq 0,$$
$$L(0) = \frac{1 + \iota}{32} - \sfr \sum_{n \in Z}
(n + \shf)\ {^\circ_\circ} e(n)e(-n) {^\circ_\circ}$$
where $\iota = 1$ if $Z = {\Bbb Z}$ and $\iota = -1$ if $Z = \BZ'$.

\proclaim{Theorem 1} The operators $L(k)$, $k\in {\Bbb Z}$, and
the identity operator, represent a $c = \shf$ Virasoro algebra $\bVir$ on
$\bCM(Z)$. In particular,
for $k,n \in {\Bbb Z}$, $m \in Z$, we have
$$[L(k),e(m)] = -(m + \shf k)\ e(m+k),$$
$$[L(k),L(n)] = (k - n)L(k + n) + {\tsize{\frac{1}{24}}}
(k^3 - k)\delta_{k,-n} 1.$$
The parity decomposition of $\bCM(Z)$ is a decomposition into two
irreducible $\bVir$-modules. The highest weight vectors in these modules
are $\bvac$, $e(-\shf)\bvac$, $\bvac'$ and $e(0)\bvac'$, whose
weights are $0$, $\shf$, ${\tsize{\frac{1}{16}}}$ and
${\tsize{\frac{1}{16}}}$, respectively.
The decomposition of $\bCM({\Bbb Z})$ into two irreducible
$\bCl({\Bbb Z})$-modules is also a
decomposition into two irreducible $\bVir$-modules. The highest weight vectors
in these modules are $\bvac'_+$ and $\bvac'_-$, whose
weights are both ${\tsize{\frac{1}{16}}}$. The operators $L(k)$ commute with
$\theta$ on $\bCM({\Bbb Z})$.
\endproclaim
\sk1

As usual we can define a positive Hermitian form $(\ ,\ )$ on $\bCM(Z)$
such that $e(m)^* = e(-m)$ and $L(k)^* = L(-k)$, where $*$ denotes
adjoint. The eigenspaces of $L(0)$ provide $\bCM(Z)$ with a grading. If
$u$ is an eigenvector for $L(0)$ write $wt(u)$ for the eigenvalue and
write $(\bCM(Z))_n$ for the $n$-eigenspace. For $u = e(-m_1)\hdots e(-m_r)
\bvac(Z) \in \bCM(Z)$ we have
$$L(0)u = \left(m_1 + \hdots + m_r + \frac{1+\iota}{32}\right) u.$$

Let ${\bold W}$ be any subspace of $\bCM(Z)$ which is a direct sum of
$L(0)$ eigenspaces, $({\bold W})_n$. The homogeneous
character of ${\bold W}$ is defined to be
$$ch({\bold W}) = \sum_n dim({\bold W})_n q^n,$$
a formal series in $q$.
The homogeneous characters of the Clifford modules are then
$$ch(\bCM(\BZ')) = \prod_{0\leq n\in {\Bbb Z}} (1 + q^{n+\shf})$$
and
$$ch(\bCM({\Bbb Z})) = 2q^{1/16} \prod_{1\leq n\in {\Bbb Z}} (1 + q^n).$$

As in the spinor construction of $D^{(1)}_n$ we construct
a vertex operator superalgebra on $\bCM(\BZ')$
and a (twisted) representation on $\bCM({\Bbb Z})$.
This is done by defining a vertex
operator $Y(v,\zeta)$ on $\bCM(\BZ')$
and on $\bCM({\Bbb Z})$ for any $v \in \bCM(\BZ')$. Actually
$Y(v,\zeta)$ is a generating function of operators
$$Y(v,\zeta) = \sum_{n\in\hf{\Bbb Z}} Y_{n+1-wt(v)}(v) \zeta^{-n-1}
= \sum_{n\in\hf{\Bbb Z}} \{v\}_n\zeta^{-n-1}$$
where
$$wt(Y_m(v)w) = wt(w) - m.$$
The definition of $Y(v,\zeta)$ on $\bCM({\Bbb Z})$
is more complicated than it is on $\bCM({\Bbb Z}+\shf)$,
but there is a common part which we denote by ${\bar Y}(v,\zeta)$.

Recall that $\bvac = \bvac(\BZ')$ and $\bvac' = \bvac({\Bbb Z})$.
On $\bCM(Z)$ define ${\bar Y}(\bvac,\zeta) =
1$ (the identity operator), and for
$0 \leq n \in {\Bbb Z}$ let
$${\bar Y}(e(-n-\shf)\bvac,\zeta) =
n!^{-1}(d/d\zeta)^n\ds{\sum_{m\in Z}} e(m) \zeta^{-m-\shf}.$$
Using the fermionic normal ordering
${^\circ_\circ} e(n_1) \cdots e(n_r) {^\circ_\circ}$
of a product of Clifford generators, for vectors of the form
$$v = e(-n_1-\shf) \cdots e(-n_r - \shf)\bvac \in \bCM(\BZ')$$
we define
$${\bar Y}(v,\zeta) =
{^\circ_\circ} {\bar Y}(e(-n_1 - \shf)\bvac,\zeta)
\cdots {\bar Y}(e(-n_r - \shf)\bvac,\zeta){^\circ_\circ}$$
and extend the definition to all $v \in \bCM(\BZ')$ by linearity. Then
$${\bar Y}_m(e(-\shf)\bvac) = e(m)$$
is a Clifford generator. Furthermore, with
$$\omega = L(-2)\bvac
= \sfr e(-\sth)e(-\shf)\bvac,$$
we have
$${\bar Y}(\omega,\zeta) =
\cases L(\zeta) &\text{on $\bCM({\Bbb Z}+\shf)$}\cr\\
L(\zeta) - {\tsize{\frac{1}{16}}}\zeta^{-2}
&\text{on $\bCM({\Bbb Z})$}\endcases$$
where $L(\zeta) = \ds{\sum_{k\in{\Bbb Z}}}L(k)\zeta^{-k-2}$
is the generating function of the Virasoro operators.

In order to define the operators $Y(v,\zeta)$ we need the additional
quadratic operator
$$\Delta(\zeta) = \sfr \sum_{0\leq m,n\in {\Bbb Z}} C_{mn} e(m+\shf)e(n+\shf)
\zeta^{-m-n-1}$$
whose definition involves the
combinatorial coefficients
$$C_{mn} = \shf\ \frac{m-n}{m+n+1} {-\shf \choose m} {-\shf \choose n}.$$
For $v \in \bCM({\Bbb Z}+\shf)$ we define
$$Y(v,\zeta) = \cases {\bar Y}(v,\zeta) &\text{on $\bCM(\BZ')$}
\cr\\
{\bar Y}(\exp(\Delta(\zeta))v,\zeta) &\text{on $\bCM({\Bbb Z})$.}
\endcases$$
Note that $\exp(\Delta(\zeta))v$ is a finite sum of vectors whose weights are
between 0 and $wt(v)$. In particular,
$$\exp(\Delta(\zeta))\omega = \omega +
{\tsize{\frac{1}{16}}}\zeta^{-2}\bvac,$$
so $Y(\omega,\zeta) = L(\zeta)$ on both $\bCM(\BZ')$ and
$\bCM({\Bbb Z})$.

\pr{Theorem 2}
We have $(\bCM(\BZ'), Y(\ ,z), \bvac, \omega)$
is a vertex operator superalgebra and
$(\bCM({\Bbb Z}), Y(\ ,z))$ is a (twisted) vertex operator superalgebra
module. \epr

\pr{Corollary 3} Let $v_i \in\bCM(\BZ')^{\alpha_i}$ for $i = 1,2$. Then for
any $r \in {\Bbb Z}$, for $m,n \in {\Bbb Z}$ on $\bCM(\BZ')$,
and for $m \in {\Bbb Z}+\shf\alpha_1$, $n \in {\Bbb Z}+\shf\alpha_2$ on
$\bCM({\Bbb Z})$, we have
$$\eqalign
{&\sum_{0 \leq i \in{\Bbb Z}}(-1)^i {r\choose i}
(\{v_1\}_{m+r-i}\{v_2\}_{n+i} - (-1)^{\alpha_1 \alpha_2 + r}
\{v_2\}_{n+r-i}\{v_1\}_{m+i})\cr
&= \sum_{0\leq k\in{\Bbb Z}} {m \choose k}
\{\{v_1\}_{r+k}v_2\}_{m+n-k}\ .\cr}$$
The sum over $k$ is finite, and we may take
$0 \leq k \leq wt(v_1) + wt(v_2) - r - 1$.  \epr

\pr{Corollary 4} For $v \in \bCM({\Bbb Z}+\shf)$, $m \in {\Bbb Z}$, on
$\bCM(Z)$ we have
$$[L(m),Y(v,z)] =
\ds{\sum_{0\leq k \in {\Bbb Z}}} {{m+1}
\choose k} z^{m+1-k} Y(L(k-1)v,z)$$
and the sum is finite.
If $L(n)v = 0$ for all $n > 0$, then we have
$$[L(m),Y(v,z)] = z^{m+1}(d/dz)Y(v,z)
+ (m+1)z^mY(L(0)v,z)$$
which means that
$$[L(m),Y_{n-m}(v)] = (-n + m\ wt(v))Y_n(v).$$
\epr

\pr{Corollary 5} For $v \in \bCM({\Bbb Z}+\shf)$, $m \in {\Bbb Z}$,
on $\bCM(Z)$ we have
$$\eqalign
{&Y(L(-m-1)v,z) = \cr
&\sum_{0\leq i \in {\Bbb Z}} {{m+i-1} \choose i}
[z^i L(-m-i-1)Y(v,z) -
(-1)^m z^{-m-i} Y(v,z)L(i-1)].\cr}$$
\epr

Let
$$\bV_0 = \bCM({\Bbb Z}+\shf)^0,\ \ \bV_2 = \bCM(\BZ')^1,$$
$$\bV_1 = \bCM({\Bbb Z})^0,\ \ \bV_3 = \bCM({\Bbb Z})^1,$$
$$\bcV = \bV_0 \oplus \bV_1 \oplus \bV_2 \oplus \bV_3,$$
$$\bV = \bV_0 \oplus \bV_2,$$
and extend the Hermitian form $(\ ,\ )$ to $\bcV$
so that $\bCM(\BZ')$ and
$\bCM({\Bbb Z})$ are orthogonal.
We give the set of subscripts $\{0,1,2,3\}$ used to index these
$\bVir$-modules (sectors) the group structure of ${\Bbb Z}_4$.
In the following table we give an orthonormal basis for the space
$(\bV_i)_{\Delta_i}$ of
vectors of minimal weight $\Delta_i$ in sector $\bV_i$.
These are vacuum vectors for $\bVir$.
$$\alignat3
&\bV_0\qquad\qquad &  u_0 &= \bvac\qquad\qquad &  \Delta_0 &= 0 \\
&\bV_1\qquad\qquad &  u_1 &= \bvac'\qquad\qquad &  \Delta_1 &= \six \\
&\bV_2\qquad\qquad &  u_2 &= \sr2 e(-\shf)\bvac\qquad\qquad &\Delta_2 &= \shf
\\
&\bV_3\qquad\qquad &  u_3 &= e(0)\bvac' \qquad\qquad & \Delta_3 &= \six
\endalignat$$

For $m,n\in {\Bbb Z}_4$, define
$$\Delta(m,n) = \Delta_m + \Delta_n - \Delta_{m+n}.$$
Then we have $\Delta(0,n) = 0$ and
$$\alignat3
\Delta(1,1) &= -\ste,\qquad& \Delta(1,2) &= \shf,\qquad& \Delta(1,3) &=\sei,\\
\Delta(2,2) &= 1,\qquad& \Delta(2,3)&=\shf,\qquad& \Delta(3,3) &= -\ste.
\endalignat$$
For $m,n,p\in {\Bbb Z}_4$, define the totally symmetric function
$$\eqalign{
\Delta(m,n,p) &= \Delta(m,n) + \Delta(m,p) - \Delta(m,n+p)\cr
&= \Delta_m + \Delta_n + \Delta_p - \Delta_{m+n} - \Delta_{m+p} - \Delta_{n+p}
+ \Delta_{m+n+p}.\cr}$$
Then we have $\Delta(0,n,p) = 0$ and
$$\alignat3
\Delta(1,1,1)&= -\fif,\quad&\Delta(1,1,2)&= 0,\quad&\Delta(1,1,3)&= -\sfr,\\
\Delta(1,2,2)&= 1,\quad&\Delta(1,2,3)&= 1,\quad&\Delta(1,3,3)&= -\sfr,\\
\Delta(2,2,2)&= 2,\quad&\Delta(2,2,3)&= 1,\quad&\Delta(2,3,3)&= 0,\quad
\Delta(3,3,3) = -\fif.\endalignat$$

\sk1
\noindent{\bf{3. Intertwining Operators}}
\sk1

We wish to extend the definition of vertex operators so that $Y(v_1,z)v_2$ is
defined for all $v_1,v_2\in \bcV$. The new operators we need to define, when
$v_1\in \bV_1\oplus\bV_3$ are called ``intertwining operators''. If $v_i\in
\bV_{n_i}$ then our basic assumptions are that the components of $Y(v_1,z)v_2$
are in $\bV_{n_1+n_2}$, that
$$[L(-1),Y(v_1,z)] = \frac{d}{dz} Y(v_1,z)$$
and that Corollaries 4 and 5 are valid for all $v\in \bcV$. Let us see to
what extent these assumptions determine the intertwiners. Since the form
$(\cdot,\cdot)$ on $\bcV$ is nondegenerate, $Y(v_1,z)v_2$ is determined if
the one-point function $(Y(v_1,z)v_2,v_3)$ is known for all $v_3\in\bcV$.
Suppose $v_i\in\bV_{n_i}$ for $1\leq i\leq 3$, $L(0)v_i = \lambda_i v_i$ and
$n_1 + n_2 = n_3$ (otherwise the one-point function is zero). For any
$0<m\in{\Bbb Z}$ we have
$$\eqalign{
&(Y(v_1,z)v_2,L(-m)v_3) = (L(m)Y(v_1,z)v_2,v_3) \cr
&= (Y(v_1,z)L(m)v_2,v_3) + ([L(m),Y(v_1,z)]v_2,v_3) \cr
&= (Y(v_1,z)L(m)v_2,v_3) + \sum_{0\leq k\in{\Bbb Z}}
{{m+1}\choose k} z^{m+1-k} (Y(L(k-1)v_1,z)v_2,v_3) \cr
&= (Y(v_1,z)L(m)v_2,v_3) + z^{m+1}\frac{d}{dz}(Y(v_1,z)v_2,v_3) \cr
&+ (m+1)z^m \lambda_1 (Y(v_1,z)v_2,v_3) \cr
&+\sum_{1<k\in{\Bbb Z}}{m+1\choose k}z^{m+1-k}(Y(L(k-1)v_1,z)v_2,v_3)\cr}$$
where we have used
$$Y(L(-1)v_1,z) = [L(-1),Y(v_1,z)] = \frac{d}{dz}Y(v_1,z)$$
which comes from Corollary 5 with $m = 0$, and our assumptions.
This shows that $(Y(v_1,z)v_2,L(-m)v_3)$ is determined by the one-point
functions $(Y(v'_1,z)v'_2,v_3)$ where $wt(v'_1) \leq wt(v_1)$ and $wt(v'_2)
\leq wt(v_2)$. This reduces the problem to the case when $v_3$ has minimal
weight in its sector, that is, when $v_3$ is a vacuum vector for $\bVir$.

Assuming that $L(k)v_3 = 0$ for all $0 < k \in {\Bbb Z}$, for $0\leq m\in
{\Bbb Z}$ we have
$$\eqalign
{&(Y(L(-m-1)v_1,z)v_2,v_3) = \cr
&\sum_{0\leq i\in{\Bbb Z}} {{m+i-1} \choose i}
[z^i (L(-m-i-1)Y(v_1,z)v_2,v_3) \cr
&\qquad -(-1)^m z^{-m-i} (Y(v_1,z)L(i-1)v_2,v_3)]\cr
&= \sum_{0\leq i\in{\Bbb Z}} {{m+i-1} \choose i}
[z^i (Y(v_1,z)v_2,L(m+i+1)v_3) \cr
&\qquad -(-1)^m z^{-m-i} (Y(v_1,z)L(i-1)v_2,v_3)].\cr}$$
But since $m\geq 0$ and $i\geq 0$, $L(m+i+1)v_3 = 0$ by assumption, so the
above equals
$$\eqalign{
&- (-1)^m \sum_{1<i\in{\Bbb Z}} {{m+i-1} \choose i}
z^{-m-i} (Y(v_1,z)L(i-1)v_2,v_3)\cr
&- (-1)^m z^{-m} (Y(v_1,z)L(-1)v_2,v_3)
- (-1)^m z^{-m-1}\lambda_2 (Y(v_1,z)v_2,v_3).\cr}$$
Now we use the fact that
$$\eqalign{
(Y(v_1,z)L(-1)v_2,v_3)
&= (L(-1)Y(v_1,z)v_2,v_3) - ([L(-1),Y(v_1,z)]v_2,v_3) \cr
&= (Y(v_1,z)v_2,L(1)v_3) - \frac{d}{dz} (Y(v_1,z)v_2,v_3)\cr
&= - \frac{d}{dz} (Y(v_1,z)v_2,v_3).\cr}$$
So $(Y(L(-m-1)v_1,z)v_2,v_3)$ for $m\geq 0$ and $v_3$ a vacuum vector is
determined by the one-point functions $(Y(v_1,z)v'_2,v_3)$ for $wt(v'_2)
\leq wt(v_2)$. Thus we are reduced to the case when $v_1$ and $v_3$ are
vacuum vectors.

With that assumption, for $0 < m\in {\Bbb Z}$, we have
$$\eqalign{
&(Y(v_1,z)L(-m)v_2,v_3) =
(L(-m)Y(v_1,z)v_2,v_3) - ([L(-m),Y(v_1,z)]v_2,v_3) \cr
&(Y(v_1,z)v_2,L(m)v_3) -
\sum_{0\leq k \in {\Bbb Z}} {{-m+1}\choose k}
z^{-m+1-k} (Y(L(k-1)v_1,z)v_2,v_3) \cr
&= - z^{-m+1} (Y(L(-1)v_1,z)v_2,v_3) -
(-m+1) z^{-m} (Y(L(0)v_1,z)v_2,v_3) \cr
&= - z^{-m+1} \frac{d}{dz} (Y(v_1,z)v_2,v_3) +
(m-1) z^{-m} \lambda_1 (Y(v_1,z)v_2,v_3) \cr}$$
showing that  $(Y(v_1,z)L(-m)v_2,v_3)$ is determined by
$(Y(v_1,z)v_2,v_3)$. Thus, we have reduced the general one-point function to
the special case when $v_1$, $v_2$ and $v_3$ are vacuum vectors. These
complex numbers are called the structure constants, and we will see if they
can be consistently determined by conditions we want for the two-point
functions.
\sk1

\noindent{\bf{4. Two-Point Functions and Hypergeometric Differential
Equations}}
\sk1

For $1\leq i \leq 4$ let $v_i \in \bV_{n_i}$ with $wt(v_i) = |v_i| = N_i +
\Delta_{n_i}$ for $0\leq N_i \in {\Bbb Z}$ and suppose $n_1 + n_2 + n_3 =
n_4$. Let
$$G(v_1,v_2,v_3,v_4;z_1,z_2) = (Y(v_1,z_1)Y(v_2,z_2)v_3,v_4),$$
$$H(v_1,v_2,v_3,v_4;z,z_2) = (Y(Y(v_1,z)v_2,z_2)v_3,v_4).$$
Then we have
$$\eqalign{
&G(v_1,v_2,v_3,v_4;z_1,z_2) = \cr
&z_1^{N_4-N_1-\Delta(n_1,n_2+n_3)} z_2^{-N_2-N_3-\Delta(n_2,n_3)}
\sum_{0\leq k\in{\Bbb Z}} \left(\frac{z_2}{z_1}\right)^k
\Phi_k(v_1,v_2,v_3,v_4)\cr}$$
where
$$\Phi_k = \Phi_k(v_1,v_2,v_3,v_4)
= (Y_{k + \Delta_{n_2+n_3} - |v_4|} (v_1)
\,Y_{-k - \Delta_{n_2+n_3} + |v_3|} (v_2) v_3 , v_4)$$
and
$$\eqalign{
&H(v_1,v_2,v_3,v_4;z,z_2) = \cr
&z^{-N_1-N_2-\Delta(n_1,n_2)}
z_2^{N_4-N_3-\Delta(n_1+n_2,n_3)}
\sum_{0\leq k\in{\Bbb Z}} \left(\frac{z}{z_2}\right)^k
\Psi_k(v_1,v_2,v_3,v_4)\cr}$$
where
$$\Psi_k = \Psi_k(v_1,v_2,v_3,v_4)
= (Y_{|v_3| - |v_4|}
(Y_{-k - \Delta_{n_1+n_2} +|v_2|} (v_1) v_2) v_3 , v_4).$$

\pr{Definition 6} For $m,n,p \in {\Bbb Z}_4$ let
$$A_{mn} = (Y_{\Delta_n - \Delta_{m+n}}(u_m)u_n, u_{m+n}),$$
$$K_{mnp} = \Phi_0(u_m,u_n,u_p,u_{m+n+p}) = A_{np}A_{m,n+p},$$
$$M_{mnp} = \Psi_0(u_m,u_n,u_p,u_{m+n+p}) = A_{mn}A_{m+n,p}.$$
\epr

\pr{Lemma 7} Let $v \in \bV_i$, $0\leq i\leq 3$, be of weight $\Delta_i$.
Then we have $L(-1)^2 v = \gamma L(-2)v$ where $\gamma$ is given by the
following table:
$$\alignat5
i&\qquad&0&\qquad&1&\qquad&2&\qquad&3&\qquad\\
\gamma&\qquad& 0&\qquad&\stf&\qquad&\sft&\qquad&\stf&\qquad\endalignat$$
\epr

\pr{Theorem 8} For $1\leq i \leq 4$ let $v_i \in \bV_{n_i}$ be vacuum
vectors for $\bVir$. For $i = 1,2$ let $Y_i = Y(v_i,z_i)$, suppose that
$L(-1)^2 v_3 = \gamma L(-2)v_3$ and let
$$G = G(v_1,v_2,v_3,v_4;z_1,z_2) = (Y_1Y_2 v_3,v_4).$$
Then $G$ satisfies the partial differential equation
$$(\partial_1 + \partial_2)^2 G + \gamma(z_1^{-1}\partial_1 +
z_2^{-1}\partial_2)G - \gamma(z_1^{-2}\Delta_{n_1} + z_2^{-2}\Delta_{n_2})G
= 0.$$ \epr

\demo{Proof} We have
$$\eqalign{
(Y_1Y_2L(-1)^2v_3,v_4) &= (Y_1 L(-1) Y_2 L(-1)v_3,v_4) - (Y_1[L(-1),Y_2]
L(-1)v_3,v_4) \cr
&= (L(-1)Y_1Y_2 L(-1)v_3,v_4) - ([L(-1),Y_1]Y_2 L(-1)v_3,v_4) \cr
&\quad - \partial_2 (Y_1Y_2L(-1)v_3,v_4) \cr
&= -(\partial_1 + \partial_2) (Y_1Y_2L(-1)v_3,v_4) \cr
&= (\partial_1 + \partial_2)^2 (Y_1Y_2v_3,v_4) \cr}$$
and
$$\eqalign{
(Y_1Y_2L(-2)v_3,v_4) &= (Y_1 L(-2) Y_2v_3,v_4) - (Y_1[L(-2),Y_2]v_3,v_4)\cr
&= - ([L(-2),Y_1]Y_2v_3,v_4) - (Y_1[L(-2),Y_2]v_3,v_4)\cr
&= - (z_1^{-1}\partial_1 - z_1^{-2}\Delta_{n_1}) (Y_1Y_2v_3,v_4)\cr
&\quad - (z_2^{-1}\partial_2 - z_2^{-2}\Delta_{n_2}) (Y_1Y_2v_3,v_4) \cr}$$
so the relation
$$(Y_1Y_2L(-1)^2v_3,v_4) = \gamma (Y_1Y_2L(-2)v_3,v_4)$$
gives the result.$\hfill\bk$ \enddemo

Let $v_i = u_{n_i}$ for $1\leq i\leq 4$, and suppose $n_4 = n_1 + n_2 + n_3$.
If we write
$$G_{n_1n_2n_3}(z_1,z_2) = K_{n_1n_2n_3} z_1^{-A} z_2^{-B}
\left(1 - \frac{z_2}{z_1}\right)^{-C} F(z_2/z_1)$$
for
$$A = \Delta(n_1,n_2+n_3),\qquad B = \Delta(n_2,n_3),\qquad C =
\Delta(n_1,n_2)$$
then the differential equation for $G$ becomes a differential equation for
$F$. Letting $x = z_2/z_1$ and using the fact that
$$\Delta(n,n_3)(\Delta(n,n_3) + 1) = \gamma(\Delta(n,n_3) + \Delta_n)$$
for any $n = 0,1,2,3$, we get the ordinary differential equation
$$x(1-x)F'' +[(\gamma-2B)+(2A+2-2C-\gamma)x]F' +[2AB-2BC+\gamma C]F=0$$
for $F$. The hypergeometric function
$$_2F_1(a,b,c;z) = \sum_{n=0}^\infty \frac{(a)_n\ (b)_n}{n!\ (c)_n} z^n$$
where $(a)_n = a(a+1)\hdots (a+n-1)$, is a solution to the hypergeometric
differential equation
$$x(1-x)w'' + (c - (a+b+1)x)w' - abw = 0.$$
Therefore, we find that
$$G_{n_1n_2n_3}(z_1,z_2) = K_{n_1n_2n_3} z_1^{-A} z_2^{-B}
\left(1-\frac{z_2}{z_1}\right)^{-C}\ {_2F_1}(a,b,c;z_2/z_1)$$
where
$$c = \gamma - 2B, \qquad
ab = (2B - \gamma)C - 2AB, \qquad
a + b = 2A - 2C - \gamma + 1.$$
In fact, we find that $a = -\Delta(n_1,n_2,n_3)$.

\pr{Theorem 9} For $1\leq i \leq 4$ let $v_i \in \bV_{n_i}$ be vacuum
vectors for $\bVir$. Suppose that
$L(-1)^2 v_3 = \gamma L(-2)v_3$ and let
$$H = H(v_1,v_2,v_3,v_4;z,z_2) = (Y(Y(v_1,z)v_2,z_2) v_3,v_4).$$
Then $H$ satisfies the partial differential equation
$$\partial_2^2 H + \gamma(z_2^{-1}\partial_2 - (z_2 + z)^{-1} z_2^{-1}
z \partial)H - \gamma((z_2+z)^{-2}\Delta_{n_1} + z_2^{-2}\Delta_{n_2})H
= 0.$$ \epr

Let $v_i = u_{n_i}$ for $1\leq i\leq 4$, and suppose $n_4 = n_1 + n_2 + n_3$.
If we write
$$H_{n_1n_2n_3}(z,z_2) = M_{n_1n_2n_3} z_2^{-A'} z^{-B'}
\left(1 - \frac{z}{z_2}\right)^{-C'} F(z/z_2)$$
for
$$A' = \Delta(n_1+n_2,n_3),\qquad B' = \Delta(n_1,n_2),\qquad C' =
\Delta(n_1,n_3)$$
then the differential equation for $H$ becomes a differential equation for
$F$. Letting $x = z/z_2$ and using the facts that
$$\Delta(n,n_3)(\Delta(n,n_3) + 1) = \gamma(\Delta(n,n_3) + \Delta_n)$$
for any $n = 0,1,2,3$, and
$$(A'-C')(A'+C'+1-\gamma) = \gamma(\Delta_{n_2} - B')$$
we get the ordinary differential equation
$$x(1+x)F'' +[(2A'+2-2\gamma)+(2A'+2-2C'-\gamma)x]F'-[2(A'-C')C'+\gamma
B']F=0$$
for $F$. Therefore, we find that
$$H_{n_1n_2n_3}(z,z_2) = M_{n_1n_2n_3} z_2^{-A'}z^{-B'}
\left(1 -\frac{z}{z_2}\right)^{-C'} \ {_2F_1}(a',b',c';-z/z_2)$$
where
$$c' = 2(A' + 1 - \gamma), \qquad
a'b' = 2(C'-A')C' - \gamma B', \qquad
a' + b' = 2A' - 2C' - \gamma + 1.$$
In fact, we find that $a' = a = -\Delta(n_1,n_2,n_3)$.

For $n_1,n_2 \in \{1,3\}$ we give the values for
$a,b,c$, $a',b',c'$, $A,B,C$, $A',B',C'$ as $n_3 = 0,1,2,3$
in tables at the end of the paper.
\sk1

\noindent{\bf{5. Transformation Formulas and Rationalization}}
\sk1

The following Lemmas are well-known results in the theory of hypergeometric
functions. The formula in Lemma 10 can easily be obtained from formula (2) on
page 92 of \cite{L}, and the formulas in Lemma 11 are the formulas
numbered (1.4.5) and (1.4.13) on page 14 of \cite{S}.
\sk1

\pr{Lemma 10} (Kummer's Quadratic Transformation Formula) For $|4z| < |1-z|^2$
we have
$${}_2F_1(a,b,1+a-b;z) = (1-z)^{-a}\ {}_2F_1(\shf a, \shf+\shf a - b,1+a-b;
-4z/(1-z)^2).$$
\epr

\pr{Lemma 11} (Gauss Recurrence Relations for Contiguous Functions) For $a$ and
$b$ nonzero, we have
$$(c-a-1)\ {}_2F_1(a,b,c;z) + a\ {}_2F_1(a+1,b,c;z) = (c-1)\
{}_2F_1(a,b,c-1;z),$$
$$c(1-z)\ {}_2F_1(a,b,c;z) + (c-a)z\ {}_2F_1(a,b,c+1;z) =
c\ {}_2F_1(a,b-1,c;z).$$
\epr

\pr{Corollary 12} For $|4z| < |1-z|^2$ we have
$${}_2F_1(\sfr,\stf,\sth; -4z/(1-z)^2) = (1 - z)^\shf,$$
$${}_2F_1(-\sfr,\sfr,\shf; -4z/(1-z)^2) = (1 - z)^{-\shf},$$
$${}_2F_1(\fif,\stf,\sth; -4z/(1-z)^2) = \frac{(1 - z)^\sth}{1+z},$$
$${}_2F_1(\sfr,\stf,\shf; -4z/(1-z)^2) = \frac{(1 - z)^\shf}{1+z}.$$
\epr

\demo{Proof} The first two equations follow from Kummer's quadratic
transformation formula by taking $a = \shf$ or $a = -\shf$, $b = 0$.
The last two equations follow from the Gauss recurrence relations by taking
$a = \sfr$, $b = \stf$, $c = \sth$ or $c = \shf$. \enddemo

We wish to rationalize the correlation functions so that we can apply the
usual contour integration techniques to obtain algebraic relations for the
vertex operators. To do this we will use substitutions which
give four sheeted coverings of the $t$-plane, where $t = z_2/z_1$.
The functions $G$ and $H$ have possible poles at $t = 0$, $t = 1$ and
$t = \infty$, as well as various cuts, but the four
sheeted coverings are branched at these points, and each of these three
points has only two points lying above it. First, in $G_{n_1n_2n_3}(z_1,z_2)$,
we use the substitution
$$t^{1/2} = \left(\frac{z_2}{z_1}\right)^\shf = \frac{2x_0}{1+x_0^2} =
\frac{2x_\infty}{1+x_\infty^2}.$$
In $G_{n_2n_1n_3}(z_2,z_1)$ we use the substitution
$$t^{-1/2} = \left(\frac{z_1}{z_2}\right)^\shf = \frac{2x_\bi}{1+x_\bi^2} =
\frac{2x_{-\bi}}{1+x_{-\bi}^2}.$$
In $H_{n_1n_2n_3}(z,z_2)$ we use the substitution
$$(t^{-1}-1)^{1/2} = \left(\frac{z_1 - z_2}{z_2}\right)^\shf
= \frac{2x_1}{1-x_1^2} = \frac{2x_{-1}}{1-x_{-1}^2}.$$
In this notation, for $\alpha\in\{0,\infty,\bi,-\bi,1,-1\}$, the variable
$x_\alpha$ is local at the point $\alpha$ on the four sheeted cover. These
points are all related to each other by linear fractional transformations,
and we let $x$ be a global variable on the genus zero covering. Below we will
discuss the regions of absolute convergence after the substitutions
have been made. Let
$$\cR_x = {\Bbb C}[x,x^{-1}(x^4-1)^{-1}]$$
be the ring of rational functions in $x$ with possible poles at those six
points. Note that if $f(x) \in \cR_x$ and $\mu(x)$ is any of the linear
fractional transformations $I(x) = x$,
$$A(x) = \frac{1}{x}, \  B(x) = \frac{x+\bi}{\bi x+1}, \
C(x) = \frac{\bi x+1}{x+\bi}, \
D(x) = \frac{x-1}{x+1} \ \hbox{or}\  E(x) = \frac{1+x}{1-x}$$
then $f(\mu(x)) \in \cR_x$. Also note that $\cR_x$ is preserved by the
operator $d/dx$.

Note that
$$1 - \frac{z_2}{z_1} = 1 - t = \frac{(1-x_0^2)^2}{(1+x_0^2)^2} =
\frac{(1-x_0^2)^4}{(1-x_0^4)^2}$$
and
$$1 + \frac{z}{z_2} = t^{-1} = \frac{(1+x_1^2)^2}{(1-x_1^2)^2} =
\frac{(1+x_1^2)^4}{(1-x_1^4)^2}.$$
The $S_3$ permutation group acting on the three points $t = 0,1,\infty$, lifts
to an $S_4$ permutation group acting on the six points $0,\infty,\bi,-\bi,
1,-1$. For $a\in\{1,-1,\bi,-\bi\}$ let $F_a(x) = ax$. The group of 24 linear
fractional transformations giving these permutations is generated by $B(x)$
and $D(x)$, each of which has order 4. It consists of the compositions of the
four functions $F_a$, with the six functions $I$, $A$, $B$, $C$, $D$ and $E$.
One has the relations $BDB^{-1} = F_\bi$, $DBD^{-1} = F_{-\bi}$,
$F_{-1}DF_{-1} = D^{-1}$ and $F_{-\bi}DF_{\bi} = B$. It is easy to check that
the correspondence $B \lra (1,2,3,4)$, $D \lra (1,3,2,4)$ and $F_\bi \lra
(1,2,4,3)$ determines an isomorphism between this group of 24 transformations
and the permutation group $S_4$.
In this paper we will not use all 24 transformations, but we will just choose
enough to relate local variables at each of the six points. However, we
believe that it will be necessary to use all 24 transformations in order to
fully understand the algebraic structure of the $2\times 2$ $B$ matrices
defined just before Theorem 15. That will be a subject for future
investigation.

We can choose local variables
$x_\alpha$ such that
$$x_\infty = \frac{1}{x_0},\qquad x_\bi = \frac{1}{x_{-\bi}},\qquad
x_1 = \frac{-1}{x_{-1}}.$$
The relations between $x_0$ and $x_\alpha$ are determined,
up to some sign choices, by their relationship to $t$. We make the choices
$$x_0 = \frac{x_\bi + \bi}{\bi x_\bi + 1} \quad\hbox{so}\quad
x_0 = \frac{- x_{-\bi} + \bi}{\bi x_{-\bi} - 1}, \qquad
x_0 = \frac{x_1 + 1}{- x_1 + 1} \quad\hbox{so}\quad
x_0 = \frac{x_{-1} - 1}{x_{-1} + 1},$$
so that
$$x_\bi = \frac{- x_0 + \bi}{\bi x_0 - 1} \quad\hbox{and}\quad
x_1 = \frac{x_0 - 1}{x_0 + 1}.$$

Until further notice we assume that $n_1,n_2,n_3 \in \{1,3\}$.

\pr{Theorem 13} For $\alpha = 0$ or $\alpha = \infty$,
after the substitution $(z_2/z_1)^{1/2} = 2x_\alpha/(1+x_\alpha^2)$,
the series
$z_2^\sei \ G_{n_1n_2n_3}(z_1,z_2)$ converges absolutely to a function
of $x_\alpha$ in the domain $|x_\alpha| < \sqrt{3 - 2\sqrt{2}}$ and
in that domain we have
$$(1-x_\alpha^4)^{\sfr} z_2^\sei\ G_{n_1n_2n_3}(z_1,z_2) =
\cases 2x_\alpha K_{n_1n_2n_3} &\text{if $n_2 = n_3$}\cr\\
K_{n_1n_2n_3} &\text{if $n_2 \ne n_3$}\endcases$$

For $\alpha = \bi$ or $\alpha = -\bi$ the above assertions with $z_1$ and
$z_2$ switched and with $n_1$ and $n_2$ switched, are true.
\epr

\demo{Proof} In the expression for $G_{n_1n_2n_3}(z_1,z_2)$ obtained after
Theorem 8 we see it as a series in $t = z_2/z_1$ which converges absolutely
for $|t| < 1$. But in order to use Corollary 12 with $z = -x_\alpha^2$ we need
$\frac{|2x_\alpha|}{|1+x_\alpha^2|} < 1$. Using polar coordinates
$x_\alpha = r e^{\bi\theta}$ this condition is equivalent to
$4r^2 < 1 + 2r^2 \cos(2\theta) + r^4$. This is certainly true when
$4r^2 < 1 - 2r^2 + r^4$, that is, when $0 < r^4 - 6r^2 + 1$. The parabola
$y = x^2 - 6x + 1$ is positive for $x < 3 - 2\sqrt{2}$ so with
$0 < r < \sqrt{3 - 2\sqrt{2}}$ we are guaranteed to have the desired
condition.

Using the values of $A,B,C,a,b,c$ in the tables, and Corollary 12, after
some algebra one gets the explicit formula stated in the theorem.
$\hfill\bk$ \enddemo

\pr{Theorem 14} For $\alpha = 1$ or $\alpha = -1$,
after the substitution $(z/z_1)^{1/2} = 2x_\alpha/(1-x_\alpha^2)$,
the series
$z^\sei \ H_{n_1n_2n_3}(z,z_2)$ converges absolutely to a function
of $x_\alpha$ in the domain $|x_\alpha| < \sqrt{3 - 2\sqrt{2}}$
and in that domain we have
$$(1-x_\alpha^4)^{\sfr} z^\sei\ H_{n_1n_2n_3}(z,z_2) =
\cases 2x_\alpha M_{n_1n_2n_3} &\text{if $n_1 = n_2$}\cr\\
M_{n_1n_2n_3} &\text{if $n_1 \ne n_2$}\endcases$$ \epr

\demo{Proof} In the expression for $H_{n_1n_2n_3}(z,z_2)$ obtained after
Theorem 9 we see it as a series in $t^{-1}-1 = z/z_2$ which converges
absolutely
for $|z/z_2| < 1$. But in order to use Corollary 12 with $z = x_\alpha^2$
we need $\frac{|2x_\alpha|}{|1-x_\alpha^2|} < 1$. Using polar coordinates
$x_\alpha = r e^{\bi\theta}$ this condition is equivalent to
$4r^2 < 1 - 2r^2 \cos(2\theta) + r^4$. This is certainly true when
$4r^2 < 1 - 2r^2 + r^4$. So the analysis proceeds just as in Theorem 13.

Using the values of $A',B',C',a',b',c'$ in the tables, and Corollary 12, after
some algebra one gets the explicit formula stated in the theorem.
$\hfill\bk$ \enddemo

In order to get the most general form of the Jacobi identity, we will
eventually apply the Cauchy residue theorem to matrix valued differential
forms on the four sheeted covering. Let
$$[G(z_1,z_2)] =
\bmatrix G_{111}(z_1,z_2) &G_{311}(z_1,z_2) &G_{333}(z_1,z_2) &G_{133}(z_1,z_2)
\\
G_{331}(z_1,z_2) &G_{131}(z_1,z_2) &G_{113}(z_1,z_2) &G_{313}(z_1,z_2)
\endbmatrix,$$
$$[G(z_2,z_1)] =
\bmatrix G_{111}(z_2,z_1) &G_{311}(z_2,z_1) &G_{333}(z_2,z_1) &G_{133}(z_2,z_1)
\\
G_{331}(z_2,z_1) &G_{131}(z_2,z_1) &G_{113}(z_2,z_1) &G_{313}(z_2,z_1)
\endbmatrix$$
and
$$[H(z,z_2)] =
\bmatrix H_{111}(z,z_2) &H_{331}(z,z_2) &H_{113}(z,z_2) &H_{333}(z,z_2)
\\
H_{311}(z,z_2) &H_{131}(z,z_2) &H_{313}(z,z_2) &H_{133}(z,z_2)
\endbmatrix.$$
Note that in the matrix $[G(z_1,z_2)] = [G_{n_1n_2n_3}(z_1,z_2)]$
the first row has $n_2 = n_3$,
the second row has $n_2 \ne n_3$, while each column has the same $n_3$ but
different values of $n_1$. The $[G(z_2,z_1)]$ matrix has the same pattern
of subscripts as the $[G(z_1,z_2)]$ matrix, but the variables $z_1$ and
$z_2$ are switched. In the matrix $[H(z,z_2)]$ the first row has $n_1 =
n_2$, the second row has $n_1 \ne n_2$, while each column has the same $n_2$
and the same $n_3$. Then, for $\alpha = 0$ or $\alpha = \infty$, we have
$$(1-x_\alpha^4)^{\sfr} z_2^\sei\ [G(z_1,z_2)]_\alpha
= \bmatrix 2K_{111}x_\alpha &2K_{311}x_\alpha &2K_{333}x_\alpha
&2K_{133}x_\alpha \\ K_{331} &K_{131} &K_{113} &K_{313} \endbmatrix\ ,$$
for $\alpha = \bi$ or $\alpha = -\bi$, we have
$$(1-x_\alpha^4)^{\sfr} z_1^\sei\ [G(z_2,z_1)]_\alpha
= \bmatrix 2K_{111}x_\alpha &2K_{311}x_\alpha &2K_{333}x_\alpha
&2K_{133}x_\alpha \\ K_{331} &K_{131} &K_{113} &K_{313} \endbmatrix\ ,$$
and for $\alpha = 1$ or $\alpha = -1$, we have
$$(1-x_\alpha^4)^{\sfr} z^\sei\ [H(z,z_2)]_\alpha
= \bmatrix 2M_{111}x_\alpha &2M_{331}x_\alpha &2M_{113}x_\alpha
&2M_{333}x_\alpha \\ M_{311} &M_{131} &M_{313} &M_{133} \endbmatrix\ .$$
These equalities mean that the left sides are series which, after an
appropriate substitution, converge
absolutely in a small disk around the appropriate $x_\alpha$ to the globally
defined matrix valued functions given on the right sides.

For any $f(x)\in\cR_x$ we can use linear fractional transformations to
express the globally defined matrix valued differential form
$$\bmatrix 2K_{111}x &2K_{311}x &2K_{333}x &2K_{133}x \\
K_{331} &K_{131} &K_{113} &K_{313} \endbmatrix f(x) dx$$
in terms of the appropriate local coordinate variable $x_\alpha$
at each of the six possible poles. The residue at each pole is then easily
found. By the residue theorem, the
sum of all the residues is zero, giving us a relation among the correlation
functions. That relation is the generalization of the Jacobi Identity which
we seek. The residue at $x = 0$ can be found immediately from the
global expression. To find the residue at $x = \infty$, use
$x_\infty = x_0^{-1}$ and find the residue at $x_\infty = 0$. Leaving off
the function $f(x)$ and the differential $dx$ for the moment, we have
$$\eqalign{
&\bmatrix 2K_{111}x_0 &2K_{311}x_0 &2K_{333}x_0 &2K_{133}x_0 \\
K_{331} &K_{131} &K_{113} &K_{313} \endbmatrix \cr
= &\bmatrix 2K_{111}x_\infty^{-1} &2K_{311}x_\infty^{-1}
&2K_{333}x_\infty^{-1} &2K_{133}x_\infty^{-1} \\
K_{331} &K_{131} &K_{113} &K_{313} \endbmatrix \cr
= &\frac{1}{x_\infty} \bmatrix 2K_{111} &2K_{311} &2K_{333} &2K_{133} \\
K_{331}x_\infty &K_{131}x_\infty &K_{113}x_\infty &K_{313}x_\infty
\endbmatrix \cr
= &\frac{1}{x_\infty} \bmatrix 0 &\frac{2K_{111}}{K_{331}} \\
\frac{K_{331}}{2K_{111}} &0 \endbmatrix \
\bmatrix 2K_{111}x_\infty &2K_{311}x_\infty &2K_{333}x_\infty &2K_{133}x_\infty
\\ K_{331} &K_{131} &K_{113} &K_{313} \endbmatrix \cr
&= (1-x_\infty^4)^{\sfr} z_2^\sei\ \frac{1}{x_\infty} \
\bmatrix 0 &\frac{2K_{111}}{K_{331}} \\ \frac{K_{331}}{2K_{111}} &0
\endbmatrix \  [G(z_1,z_2)]_\infty \cr}$$
if and only if the following conditions are consistent:
$$K_{111}K_{131} = K_{331}K_{311},\qquad K_{111}K_{113} = K_{331}K_{333},
\qquad K_{111}K_{313} = K_{331}K_{133}.$$

To find the residue at $x = \bi$ use $x_0 = (x_\bi+\bi)/(\bi x_\bi+1)$
and find the residue at $x_\bi = 0$. Without imposing any further
conditions, we have
$$\eqalign{
&\bmatrix 2K_{111}x_0 &2K_{311}x_0 &2K_{333}x_0 &2K_{133}x_0 \\
K_{331} &K_{131} &K_{113} &K_{313} \endbmatrix \cr
= &\bmatrix 2K_{111}\frac{x_\bi+\bi}{\bi x_\bi+1}
&2K_{311} \frac{x_\bi+\bi}{\bi x_\bi+1}
&2K_{333} \frac{x_\bi+\bi}{\bi x_\bi+1}
&2K_{133} \frac{x_\bi+\bi}{\bi x_\bi+1}
\\ K_{331} &K_{131} &K_{113} &K_{313} \endbmatrix \cr
= &\frac{1}{\bi x_\bi+1} \bmatrix 2K_{111} (x_\bi+\bi) &2K_{311} (x_\bi+\bi)
&2K_{333} (x_\bi+\bi) &2K_{133} (x_\bi+\bi)
\\ K_{331}(\bi x_\bi+1) &K_{131}(\bi x_\bi+1) &K_{113}(\bi x_\bi+1)
&K_{313}(\bi x_\bi+1) \endbmatrix \cr
= &\frac{1}{\bi x_\bi+1} \bmatrix 1 &\frac{2\bi K_{111}}{K_{331}} \\
\frac{\bi K_{331}}{2K_{111}} &1 \endbmatrix \
\bmatrix 2K_{111}x_\bi &2K_{311}x_\bi &2K_{333}x_\bi &2K_{133}x_\bi \\
K_{331} &K_{131} &K_{113} &K_{313} \endbmatrix \cr
&= (1-x_\bi^4)^{\sfr} z_1^\sei\ \frac{1}{\bi x_\bi+1} \
\bmatrix 1 &\frac{2\bi K_{111}}{K_{331}} \\
\frac{\bi K_{331}}{2K_{111}} &1 \endbmatrix \  [G(z_2,z_1)]_\bi.\cr}$$

To find the residue at $x=-\bi$ use $x_0=(- x_{-\bi}+\bi)/(\bi x_{-\bi}-1)$
and find the residue at $x_{-\bi} = 0$. Without imposing any new conditions,
we have
$$\eqalign{
&\bmatrix 2K_{111}x_0 &2K_{311}x_0 &2K_{333}x_0 &2K_{133}x_0 \\
K_{331} &K_{131} &K_{113} &K_{313} \endbmatrix \cr
= &\frac{-1}{\bi x_{-\bi}-\!1}\!\!\bmatrix 2K_{111} (x_{-\bi}-\bi)\!\!\!\!
&2K_{311} (x_{-\bi}-\bi)\!\!\!\! &2K_{333} (x_{-\bi}-\bi)\!\!\!\!
&2K_{133} (x_{-\bi}-\bi)
\\ K_{331}(-\bi x_{-\bi}\!+\!1)\!\!\!\! &K_{131}(-\bi x_{-\bi}\!+\!1)\!\!\!\!
&K_{113}(-\bi x_{-\bi}+\!1) \!\!\!\!
&K_{313}(-\bi x_{-\bi}+\!1) \endbmatrix \cr
= &\frac{1}{\bi x_{-\bi}-1} \bmatrix -1 &\frac{2\bi K_{111}}{K_{331}} \\
\frac{\bi K_{331}}{2K_{111}} &-1 \endbmatrix \
\bmatrix 2K_{111}x_{-\bi} &2K_{311}x_{-\bi}
&2K_{333}x_{-\bi} &2K_{133}x_{-\bi} \\
K_{331} &K_{131} &K_{113} &K_{313} \endbmatrix \cr
&= (1-x_{-\bi}^4)^{\sfr} z_1^\sei\
\frac{1}{\bi x_{-\bi}-1} \bmatrix -1 &\frac{2\bi K_{111}}{K_{331}} \\
\frac{\bi K_{331}}{2K_{111}} &-1 \endbmatrix \ [G(z_2,z_1)]_{-\bi} .\cr}$$

To find the residue at $x = -1$ use $x_0 = (x_{-1} - 1)/(x_{-1} + 1)$
and find the residue at $x_{-1} = 0$. Assuming the previous consistency
conditions are true, we have
$$\eqalign{
&\bmatrix 2K_{111}x_0 &2K_{311}x_0 &2K_{333}x_0 &2K_{133}x_0 \\
K_{331} &K_{131} &K_{113} &K_{313} \endbmatrix \cr
= &\frac{1}{x_{-1}+1}\bmatrix 2K_{111}(x_{-1}-1)\!\!\!&2K_{311}(x_{-1}-1)\!\!\!
&2K_{333} (x_{-1}-1)\!\!\! &2K_{133} (x_{-1}-1) \\ K_{331}(x_{-1}+1)\!\!\!
&K_{131}(x_{-1}+1)\!\!\!&K_{113}(x_{-1}+1)\!\!\!&K_{313}(x_{-1}+1)\endbmatrix\cr
= &\frac{1}{x_{-1}+1} \bmatrix \frac{K_{111}}{M_{111}}
&\frac{-2K_{111}}{M_{311}} \\
\frac{K_{331}}{2M_{111}} &\frac{K_{331}}{M_{311}} \endbmatrix
\bmatrix 2M_{111}x_{-1} &2M_{331}x_{-1} &2M_{113}x_{-1} &2M_{333}x_{-1} \\
M_{311} &M_{131} &M_{313} &M_{133} \endbmatrix \cr
= &(1-x_{-1}^4)^{\sfr} z^\sei\ \frac{1}{x_{-1}+1}
\bmatrix \frac{K_{111}}{M_{111}} &\frac{-2K_{111}}{M_{311}} \\
\frac{K_{331}}{2M_{111}} &\frac{K_{331}}{M_{311}} \endbmatrix \
[H(z,z_2)]_{-1} \cr}$$
if and only if the following conditions hold:
$$\frac{K_{111}}{K_{133}}=\frac{M_{111}}{M_{333}}
=\frac{M_{311}}{M_{133}},\ \
\frac{K_{111}}{K_{311}}=\frac{M_{111}}{M_{331}}
=\frac{M_{311}}{M_{131}},\ \
\frac{K_{111}}{K_{333}}=\frac{M_{111}}{M_{113}}=\frac{M_{311}}{M_{313}}.$$

To find the residue at $x = 1$ use $x_0 = (x_1 + 1)/(-x_1 + 1)$
and find the residue at $x_1 = 0$. Without imposing any new conditions, we have
$$\eqalign{
&\bmatrix 2K_{111}x_0 &2K_{311}x_0 &2K_{333}x_0 &2K_{133}x_0 \\
K_{331} &K_{131} &K_{113} &K_{313} \endbmatrix \cr
= &\frac{1}{-x_1+1} \bmatrix 2K_{111} (x_1+1)\!\! &2K_{311} (x_1+1)\!\!
&2K_{333} (x_1+1)\!\! &2K_{133} (x_1+1) \\ K_{331}(-x_1+1)\!\!
&K_{131}(-x_1+1)\!\! &K_{113}(-x_1+1)\!\! &K_{313}(-x_1+1)\endbmatrix \cr
= &\frac{1}{-x_1+1}\bmatrix\frac{K_{111}}{M_{111}} &\frac{2K_{111}}{M_{311}}\\
\frac{-K_{331}}{2M_{111}} &\frac{K_{331}}{M_{311}} \endbmatrix \
\bmatrix 2M_{111}x_1 &2M_{331}x_1 &2M_{113}x_1 &2M_{333}x_1 \\
M_{311} &M_{131} &M_{313} &M_{133} \endbmatrix \cr
&= (1-x_1^4)^{\sfr} z^\sei\ \frac{1}{-x_1+1}
\bmatrix \frac{K_{111}}{M_{111}} &\frac{2K_{111}}{M_{311}} \\
\frac{-K_{331}}{2M_{111}} &\frac{K_{331}}{M_{311}} \endbmatrix \
[H(z,z_2)]_1 .\cr}$$

Using Definition 6, the consistency conditions we have found translate into
the following conditions on the structure constants $A_{mn}$:
$$\frac{A_{30}}{A_{10}}=\frac{A_{12}}{A_{32}}=\frac{A_{11}}{A_{33}}=
\frac{A_{31}}{A_{13}},\ \
\frac{A_{30}A_{31}}{A_{10}A_{13}} = 1,\ \ A_{21}=A_{23},\ \ A_{01}=A_{03}.$$

In fact, we know quite a bit more about the structure constants $A_{mn}$. Since
$Y(u_0,z) = Y(\bvac,z) = 1$, we have $A_{0n} = 1$ for any $n$. We also want
$$Y(v,z)\bvac = e^{zL(-1)} v \in \bcV[z]$$
which implies the ``creation property''
$$lim_{z\to 0} Y(v,z)\bvac = Y_{-wt(v)}(v)\bvac = v.$$
These give $A_{m0} = 1$ for any $m$. We also have
$$A_{21} = (Y_0(u_2)u_1,u_3) = \frac{1}{\sqrt{2}} (e(0)\bvac',e(0)\bvac')
= \frac{1}{\sqrt{2}},$$
$$A_{22} = (Y_{1/2}(u_2)u_2,u_0) = \shf (e(\shf)e(-\shf)\bvac,\bvac)
= \shf \langle e,e\rangle (\bvac,\bvac) = 1,$$
$$A_{23} = (Y_0(u_2)u_3,u_1) = \frac{1}{\sqrt{2}} (e(0)e(0)\bvac',\bvac')
= \frac{1}{\sqrt{2}}.$$
We also want the ``symmetry'' condition, generalizing what is called
``skew-symmetry'' in \cite{FHL},
$$z^{\Delta(n_1,n_2)} Y(v_1,z)v_2 = (-z)^{\Delta(n_1,n_2)} e^{zL(-1)}
Y(v_2,-z)v_1$$
for $v_i \in \bV_{n_i}$. With $v_1 = u_m$ and $v_2 = u_n$, after pairing
with $u_{m+n}$, this gives $A_{mn} = A_{nm}$. Along with the above
constraints, this determines all the structure constants $A_{mn}$ except
$A_{11}$, which we normalize to $1/\sqrt{2}$, and $A_{13}$,
which we normalize to $1$. Here is a table summarizing the results.
$$\alignat5
A&_{mn}&\qquad n&=0 &\qquad n&=1 &\qquad n&=2 &\qquad n&=3 \\
m&=0 &\qquad &1 &\qquad &1 &\qquad &1 &\qquad &1 \\
m&=1 &\qquad &1 &\qquad &\frac{1}{\sqrt{2}} &\qquad &\frac{1}{\sqrt{2}}
&\qquad &1 \\
m&=2 &\qquad &1 &\qquad &\frac{1}{\sqrt{2}} &\qquad &1 &\qquad
&\frac{1}{\sqrt{2}} \\
m&=3 &\qquad &1 &\qquad &1 &\qquad &\frac{1}{\sqrt{2}} &\qquad
&\frac{1}{\sqrt{2}}
\endalignat$$

\vskip 5pt
\pr{Definition} Define the matrices
$$\eqalign{
B_0 &= \bmatrix 1 &0 \\ 0 &1\endbmatrix,\qquad
B_\infty = \bmatrix 0 &1 \\ 1 &0\endbmatrix,\qquad
B_{\bi} = \bmatrix 1 &\bi \\ \bi &1\endbmatrix,\cr
B_{-\bi} &= \bmatrix -1 &\bi \\ \bi &-1\endbmatrix,
\qquad B_{-1} = \bmatrix 1 &-1 \\ 1 &1\endbmatrix,\qquad
B_1 = \bmatrix 1 &1 \\ -1 &1\endbmatrix .\cr}$$ \epr

\pr{Theorem 15} With $A_{mn}$ as given in the above table, and with
notations as defined above, we have
$$\eqalign{
&(1-x_0^4)^{\sfr} z_2^\sei\ [G(z_1,z_2)]_0 \cr
\sim &(1-x_\infty^4)^{\sfr} z_2^\sei\ x_\infty^{-1} B_\infty\
[G(z_1,z_2)]_\infty\cr
\sim &(1-x_\bi^4)^{\sfr} z_1^\sei\ (\bi x_\bi+1)^{-1} B_{\bi}\
[G(z_2,z_1)]_\bi \cr
\sim &(1-x_{-\bi}^4)^{\sfr} z_1^\sei\ (\bi x_{-\bi}-1)^{-1}
B_{-\bi} \ [G(z_2,z_1)]_{-\bi} \cr
\sim &(1-x_{-1}^4)^{\sfr} z^\sei\ (x_{-1}+1)^{-1} B_{-1} \ [H(z,z_2)]_{-1} \cr
\sim &(1-x_1^4)^{\sfr} z^\sei\ (-x_1+1)^{-1} B_1 \ [H(z,z_2)]_1 \cr}$$
where $\sim$ means that these series converge absolutely in their
appropriate domains to the same globally defined matrix valued function
$$\bmatrix x_0 &x_0 &x_0 &x_0 \\
1 &1 &1 &1 \endbmatrix $$
on the four sheeted covering. \epr

Note that if we associate to a matrix $M = \bmatrix a&b \\ c&d\endbmatrix$
a linear fractional transformation $f(x) = \frac{ax+b}{cx+d}$
then we associate to the matrix $B_\alpha$ the linear fractional transformation
$$f_\alpha(x_\alpha) = x_0 = \frac{ax_\alpha+b}{cx_\alpha+d}.$$
\vskip 20pt

\noindent{\bf{6. Inductive formulas}}
\sk1

Now let us see how the general two-point functions are determined inductively.

\pr{Theorem 16} Let $v_i \in \bV_{n_i}$, $1\leq i \leq 4$, be eigenvectors
for $L(0)$ with $L(0)v_i =$\newline $wt(v_i)v_i = |v_i| v_i$.
For $i = 1,2$ let $Y_i = Y(v_i,z_i)$, $\partial_i = \partial/\partial z_i$,
and recall the notation
$$G(v_1,v_2,v_3,v_4;z_1,z_2) = (Y_1Y_2 v_3,v_4).$$
Then for any $k \in {\Bbb Z}$ we have
$$\eqalign{
G(v_1,v_2,v_3,L(-k)v_4;z_1,z_2) &= G(v_1,v_2,L(k)v_3,v_4;z_1,z_2) \cr
&+ (z_1^{k+1}\partial_1 + z_2^{k+1}\partial_2) G(v_1,v_2,v_3,v_4;z_1,z_2)\cr
&+ (k+1) ( z_1^k |v_1| + z_2^k |v_2|) G(v_1,v_2,v_3,v_4;z_1,z_2) \cr
&+ \sum_{i\geq 1} {{k+1}\choose{i+1}}z_1^{k-i}
G(L(i)v_1,v_2,v_3,v_4;z_1,z_2)\cr
&+ \sum_{i\geq 1} {{k+1}\choose{i+1}}z_2^{k-i}
G(v_1,L(i)v_2,v_3,v_4;z_1,z_2).\cr}$$
\epr

\demo{Proof} We have
$$\eqalign{
G(v_1,&v_2,v_3,L(-k)v_4;z_1,z_2) = (L(k)Y_1Y_2 v_3,v_4) \cr
&= ([L(k),Y_1]Y_2 v_3,v_4) + (Y_1[L(k),Y_2] v_3,v_4) + (Y_1Y_2 L(k)v_3,v_4)\cr
&= \sum_{i\geq 0} {{k+1}\choose{i}}z_1^{k+1-i} (Y(L(i-1)v_1,z_1)Y_2v_3,v_4)\cr
&+ \sum_{i\geq 0} {{k+1}\choose{i}}z_2^{k+1-i} (Y_1Y(L(i-1)v_2,z_2)v_3,v_4)\cr
&+ (Y_1Y_2 L(k)v_3,v_4)\cr}$$
giving the result after separating the $i = 0$ and the $i = 1$ terms
and reindexing. $\bk$  \enddemo

\pr{Theorem 17} Let $v_i \in \bV_{n_i}$, $1\leq i \leq 4$, be eigenvectors
for $L(0)$ with $L(0)v_i =$\newline $wt(v_i)v_i = |v_i| v_i$.
Let $\partial_1 = \partial/\partial z_1$, $\partial = \partial/\partial z$,
and recall the notation
$$H(v_1,v_2,v_3,v_4;z,z_2) = (Y(Y(v_1,z)v_2,z_2) v_3,v_4).$$
Then for any $k \in {\Bbb Z}$ we have
$$\eqalign{
&H(v_1,v_2,v_3,L(-k)v_4;z,z_2) \cr
&= H(v_1,v_2,L(k)v_3,v_4;z,z_2) \cr
&+ z_2^{k+1}\partial_2H(v_1,v_2,v_3,v_4;z,z_2)\cr
&+ (k+1)(z_2^k |v_2| + (z_2 + z)^k |v_1|) H(v_1,v_2,v_3,v_4;z,z_2) \cr
&+ [(z_2 + z)^{k+1} - z_2^{k+1}] \partial H(v_1,v_2,v_3,v_4;z,z_2)\cr
&+ \sum_{i\geq 1} {{k+1}\choose{i+1}} z_2^{k-i} H(v_1,L(i)v_2,v_3,v_4;z,z_2)\cr
&+ \sum_{j\geq 1}{{k+1}\choose{j+1}}(z_2+z)^{k-j} H(L(j)v_1,v_2,v_3,v_4;z,z_2)
.\cr}$$
\epr

\demo{Proof} For brevity, let us write
$$H = H(v_1,v_2,v_3,v_4;z,z_2),\quad
H(L(k)v_3) = H(v_1,v_2,L(k)v_3,v_4;z,z_2),$$
$$H(L(j)v_1) = H(L(j)v_1,v_2,v_3,v_4;z,z_2)$$
and
$$H(L(i)v_2) = H(v_1,L(i)v_2,v_3,v_4;z,z_2).$$
We have
$$\eqalign{
&H(v_1,v_2,v_3,L(-k)v_4;z,z_2) \cr
&= ([L(k),Y(Y(v_1,z)v_2,z_2)]v_3,v_4) +
(Y(Y(v_1,z)v_2,z_2)L(k)v_3,v_4)\cr
&= \sum_{i\geq 0} {{k+1}\choose{i}} z_2^{k+1-i}
(Y(L(i-1)Y(v_1,z)v_2,z_2) v_3,v_4) + H(L(k)v_3) \cr
&= z_2^{k+1} \partial_2 H + H(L(k)v_3) +
\sum_{i\geq 0} {{k+1}\choose{i+1}} z_2^{k-i}
(Y([L(i),Y(v_1,z)]v_2,z_2) v_3,v_4) \cr
&+ (k+1)z_2^k |v_2| H + \sum_{i\geq 1} {{k+1}\choose{i+1}} z_2^{k-i}
(Y(Y(v_1,z)L(i)v_2,z_2) v_3,v_4) \cr
&= z_2^{k+1} \partial_2 H + H(L(k)v_3) \cr
&+ \sum_{i\geq 0} {{k+1}\choose{i+1}} z_2^{k-i}
\left(z^{i+1} \partial H + (i+1)z^i |v_1| H +
\sum_{j\geq 1} {{i+1}\choose{j+1}} z^{i-j} H(L(j)v_1) \right) \cr
&+ (k+1)z_2^k |v_2| H + \sum_{i\geq 1} {{k+1}\choose{i+1}} z_2^{k-i}
H(L(i)v_2) \cr}$$
$$\eqalign{
&= z_2^{k+1} \partial_2 H + H(L(k)v_3)
+ z_2^{k+1} \sum_{i\geq 0} {{k+1}\choose{i+1}}
\left(\frac{z}{z_2}\right)^{i+1} \partial H \cr
&+ |v_1| z_2^k H \sum_{i\geq 0} {{k+1}\choose{i+1}} (i+1)
\left(\frac{z}{z_2}\right)^{i} \cr
&+ z_2^{k-j} \sum_{j\geq 1} H(L(j)v_1) \sum_{i\geq 0} {{k+1}\choose{i+1}}
{{i+1}\choose{j+1}} \left(\frac{z}{z_2}\right)^{i-j} \cr
&+ (k+1)z_2^k |v_2| H + \sum_{i\geq 1} {{k+1}\choose{i+1}} z_2^{k-i}
H(L(i)v_2) \cr}$$
which gives the result after using
$${{k+1}\choose{i+1}} (i+1) = (k+1) {k\choose i}$$
and
$$(z_2 + z)^m = z_2^m \left(1 + \frac{z}{z_2}\right)^m =
z_2^m \sum_{i\geq 0} {m\choose i} \left(\frac{z}{z_2}\right)^i.$$
$\hfill\bk$ \enddemo

For $1\leq i \leq 4$ let $v_i \in \bV_{n_i}$ with $wt(v_i) = |v_i| = N_i +
\Delta_{n_i}$ for $0\leq N_i \in {\Bbb Z}$. For $n_1,n_2,n_3\in\{1,3\}$ we
have
$$\Delta(n_1,n_2+n_3) =
\cases 0 &\text{if $n_2+n_3 = 0$}\cr\\
\shf &\text{if $n_2+n_3 = 2$}\endcases
\ \hbox{ and }\
\Delta(n_2,n_3) =
\cases \sei &\text{if $n_2+n_3 = 0$}\cr\\
-\ste &\text{if $n_2+n_3 = 2$}\endcases$$
so that
$$\Delta(n_1,n_2+n_3) + \Delta(n_2,n_3) = \sei$$
for $n_1,n_2,n_3\in\{1,3\}$. Let $N = N_1 + N_2 + N_3 - N_4$.
With $t = z_2/z_1$ and
$\Delta = \Delta(n_1,n_2+n_3)$ we have
$$G(v_1,v_2,v_3,v_4;z_1,z_2)
= t^{N_1-N_4+\Delta} z_2^{-N-\sei} \Phi(v_1,v_2,v_3,v_4;t)$$
where
$$\Phi(v_1,v_2,v_3,v_4;t) = \sum_{0\leq k\in{\Bbb Z}} t^k
\Phi_k(v_1,v_2,v_3,v_4)$$
is a power series in $t$.
Similarly, with $s = z/z_2 = t^{-1} - 1$ and
${\bDel} = \Delta(n_1+n_2,n_3)$ we have
$$H(v_1,v_2,v_3,v_4;z,z_2)
= s^{N_3-N_4+\bDel} z^{-N-\sei} \Psi(v_1,v_2,v_3,v_4;s)$$
where
$$\Psi(v_1,v_2,v_3,v_4;s) = \sum_{0\leq k\in{\Bbb Z}} s^k
\Psi_k(v_1,v_2,v_3,v_4)$$
is a power series in $s$.

We choose the relationship
$s^{1/2} = (t^{-1} - 1)^{1/2}$ such that we have
$(1-x_0^2)/2x_0 = -2x_1/(1-x_1^2)$.
We wish to rewrite the recursions for $G$ and $H$ given in the last two
theorems as recursions for the functions
$$\eqalign{
{\tPhi}(v_1,v_2,v_3,v_4;x) &= (1-x^4)^\sfr z_2^{N+\sei}
G(v_1,v_2,v_3,v_4;z_1,z_2) \cr
&= (1-x^4)^\sfr t^{N_1-N_4+\Delta} \Phi(v_1,v_2,v_3,v_4;t)\cr} $$
where $x = x_0$ or $x = x_\infty$ and $t^{1/2} = 2x/(1+x^2)$, and
$$\eqalign{
{\tPsi}(v_1,v_2,v_3,v_4;w)&=(1-w^4)^\sfr z^{N+\sei} H(v_1,v_2,v_3,v_4;z,z_2)\cr
&= (1-w^4)^\sfr s^{N_3-N_4+\bDel} \Psi(v_1,v_2,v_3,v_4;s) \cr} $$
where $w = x_1$ or $w = x_{-1}$ and $s^{1/2} = 2w/(1-w^2)$.

We will later apply the first of these recursions to the functions
$$\eqalign{
{\tPhi}(v_2,v_1,v_3,v_4;y) &= (1-y^4)^\sfr z_1^{N+\sei}
G(v_2,v_1,v_3,v_4;z_2,z_1) \cr
&= (1-y^4)^\sfr t^{-N_2+N_4-\Delta'} \Phi(v_2,v_1,v_3,v_4;t^{-1})\cr} $$
where $y = x_\bi$ or $y = x_{-\bi}$ and $\Delta' = \Delta(n_2,n_1+n_3)$.
These will be obtained from the first
kind of recursions by switching $v_1$ with $v_2$, $z_1$ with
$z_2$, and $n_1$ with $n_2$, and by replacing $t$ by $t^{-1}$.
Define the matrices
$$[\tPhi_{top}]_\alpha = (1-x_\alpha^4)^\sfr z_2^\sei
[G(z_1,z_2)]_\alpha$$
for $\alpha = 0$ or $\alpha = \infty$,
$$[\tPhi_{top}]_\alpha = (1-x_\alpha^4)^\sfr z_1^\sei
[G(z_2,z_1)]_\alpha$$
for $\alpha = \bi$ or $\alpha = -\bi$, and
$$[\tPsi_{top}]_\alpha = (1-x_\alpha^4)^\sfr z^\sei
[H(z,z_2)]_\alpha$$
for $\alpha = 1$ or $\alpha = -1$. Then our earlier work, which will give the
base cases of our inductions, can be written as
$$\eqalign{
[\tPhi_{top}]_0
&\sim x_\infty^{-1} B_\infty\ [\tPhi_{top}]_\infty \cr
\sim (\bi x_\bi+1)^{-1} B_{\bi}\ [\tPhi_{top}]_\bi
&\sim (\bi x_{-\bi}-1)^{-1} B_{-\bi}\ [\tPhi_{top}]_{-\bi} \cr
\sim (x_{-1}+1)^{-1} B_{-1}\ [\tPsi_{top}]_{-1}
&\sim (-x_1+1)^{-1} B_1 \ [\tPsi_{top}]_1 \cr}$$

The general case requires us to define $2\times 4$ matrices which are
analogs of $[\tPhi_{top}]_\alpha$ and $[\tPsi_{top}]_\alpha$ with the
``top'' vectors $u_{n_i}$ replaced by general vectors. There is no loss of
generality in assuming that these vectors $v_i$ are $L(0)$-eignevectors with
$wt(v_i) = N_i + \Delta_{n_i}$, $1\leq i\leq 4$. We must do this in such a
way that our induction formulas imply that the above top $\sim$ relations
hold for the more general matrices. This means that we need a more general
description than the pattern of subscripts defining the matrices
$[G(z_1,z_2)]$, $[G(z_2,z_1)]$ and $[H(z,z_2)]$. We will use the isomorphism
$\theta$ on $\bCM({\Bbb Z})$ defined at the beginning of this paper to give
such a description. Assume that $v_1,v_2,v_3 \in \bV_1$
and $v_4\in\bV_3$ are $L(0)$-eigenvectors with weights as given
above. In the entries of the following matrices we will write the functions
$\tPhi(v_1,v_2,v_3,v_4;x_\alpha)$ and $\tPsi(v_1,v_2,v_3,v_4;x_\alpha)$
more briefly as just
$\tPhi(v_1,v_2,v_3,v_4)$ and $\tPsi(v_1,v_2,v_3,v_4)$.
For $\alpha = 0$ or $\alpha = \infty$, define the matrix
$$\eqalign{
&[\tPhi]_\alpha = [\tPhi(v_1,v_2,v_3,v_4;x_\alpha)] = \cr
&\bmatrix \tPhi(v_1,v_2,v_3,v_4) \!\!\!\!
&\tPhi(\theta v_1,v_2,v_3,\theta v_4) \!\!\!\!
&\tPhi(\theta v_1,\theta v_2,\theta v_3,\theta v_4) \!\!\!\!
&\tPhi(v_1,\theta v_2,\theta v_3,v_4) \\
\tPhi(\theta v_1,\theta v_2,v_3,v_4)  \!\!\!\!
&\tPhi(v_1,\theta v_2,v_3,\theta v_4) \!\!\!\!
&\tPhi(v_1,v_2,\theta v_3,\theta v_4)  \!\!\!\!
&\tPhi(\theta v_1,v_2,\theta v_3,v_4)\endbmatrix.\cr}$$
For $\alpha = \bi$ or $\alpha = -\bi$ define the matrix
$$\eqalign{
&[\tPhi]_\alpha = [\tPhi(v_1,v_2,v_3,v_4;x_\alpha)] = \cr
&\bmatrix \tPhi(v_2,v_1,v_3,v_4) \!\!\!\!
&\tPhi(\theta v_2,v_1,v_3,\theta v_4) \!\!\!\!
&\tPhi(\theta v_2,\theta v_1,\theta v_3,\theta v_4)  \!\!\!\!
&\tPhi(v_2,\theta v_1,\theta v_3,v_4) \\
\tPhi(\theta v_2,\theta v_1,v_3,v_4)  \!\!\!\!
&\tPhi(v_2,\theta v_1,v_3,\theta v_4) \!\!\!\!
&\tPhi(v_2,v_1,\theta v_3,\theta v_4)  \!\!\!\!
&\tPhi(\theta v_2,v_1,\theta v_3,v_4)\endbmatrix.\cr}$$
For $\alpha = 1$ or $\alpha = -1$ define the matrix
$$\eqalign{
&[\tPsi]_\alpha = [\tPsi(v_1,v_2,v_3,v_4;x_\alpha)] = \cr
&\bmatrix \tPsi(v_1,v_2,v_3,v_4)  \!\!\!\!
&\tPsi(\theta v_1,\theta v_2,v_3,v_4) \!\!\!\!
&\tPsi(v_1,v_2,\theta v_3,\theta v_4)  \!\!\!\!
&\tPsi(\theta v_1,\theta v_2,\theta v_3,\theta v_4) \\
\tPsi(\theta v_1,v_2,v_3,\theta v_4)  \!\!\!\!
&\tPsi(v_1,\theta v_2,v_3,\theta v_4) \!\!\!\!
&\tPsi(\theta v_1,v_2,\theta v_3,v_4)  \!\!\!\!
&\tPsi(v_1,\theta v_2,\theta v_3,v_4)\endbmatrix.\cr}$$

In our inductions we will
see such matrices with one of the vectors, say $v_i$, replaced by
$L(-k)v_i$. We will denote the above matrices with such a replacement by
$[\tPhi(L(-k)v_i)]_\alpha$ and $[\tPsi(L(-k)v_i)]_\alpha$.

Let $\Phi' = \frac{\partial}{\partial t} \Phi$,
$\Psi' = \frac{\partial}{\partial s} \Psi$ and let us use notations such as
$\tPhi(L(i)v_1) = \tPhi(L(i)v_1,v_2,v_3,v_4;x)$ as we did in the proof of
Theorem 17. We have
$$\partial_1 G = \frac{N_4-N_1-\Delta}{z_1} G - \frac{z_2}{z_1^2}
t^{N_1-N_4+\Delta} z_2^{-N-\sei} \Phi'$$
$$\partial_2 G = \frac{\Delta-N_2-N_3-\sei}{z_2} G + \frac{1}{z_1}
t^{N_1-N_4+\Delta} z_2^{-N-\sei} \Phi'$$
$$\partial H = \frac{-N_1-N_2-\sei+\bDel}{z} H + \frac{1}{z_2}
s^{N_3-N_4+\bDel} z^{-N-\sei} \Psi'$$
$$\partial_2 H = \frac{N_4-N_3-\bDel}{z_2} H - \frac{z}{z_2^2}
s^{N_3-N_4+\bDel} z^{-N-\sei} \Psi'.$$

\pr{Theorem 18} Define the operator
$$\cL_x = \frac{x(1+x^2)}{2(1-x^2)} \frac{d}{dx} + \frac{x^4}{2(1-x^2)^2}
= \frac{x(1+x^2)}{2(1-x^2)} \left[\frac{d}{dx} + \frac{x^3}{1-x^4}\right].$$
Then for any $k\in {\Bbb Z}$ we have
$$\eqalign{
\tPhi(L(-k)v_4)
=\ &\tPhi(L(k)v_3) - (N + \sei)\tPhi + (1-t^{-k}) \cL_x \tPhi \cr
&+ \sum_{i\geq 0} {{k+1}\choose{i+1}} \left(t^{i-k} \tPhi(L(i)v_1)
+ \tPhi(L(i)v_2)\right). \cr}$$
\epr

\demo{Proof} For any $k\in {\Bbb Z}$ we have
$$\eqalign{
\tPhi(L(-k)v_4) &=
(1-x^4)^\sfr z_2^{N-k+\sei} G(v_1,v_2,v_3,L(-k)v_4;z_1,z_2) \cr
&= (1-x^4)^\sfr z_2^{N-k+\sei} G(v_1,v_2,L(k)v_3,v_4;z_1,z_2) \cr
&+ (1-x^4)^\sfr z_2^{N-k+\sei} (z_1^{k+1}\partial_1 + z_2^{k+1}\partial_2)
G(v_1,v_2,v_3,v_4;z_1,z_2)\cr
&+ (1-x^4)^\sfr z_2^{N-k+\sei} (k+1) ( z_1^k |v_1| + z_2^k |v_2|)
G(v_1,v_2,v_3,v_4;z_1,z_2) \cr
&+ (1-x^4)^\sfr z_2^{N-k+\sei} \sum_{i\geq 1} {{k+1}\choose{i+1}}z_1^{k-i}
G(L(i)v_1,v_2,v_3,v_4;z_1,z_2)\cr
&+ (1-x^4)^\sfr z_2^{N-k+\sei} \sum_{i\geq 1} {{k+1}\choose{i+1}}z_2^{k-i}
G(v_1,L(i)v_2,v_3,v_4;z_1,z_2) \cr
&= \tPhi(L(k)v_3) + ((N_4-N_1-\Delta)t^{-k} + (\Delta-N_2-N_3-\sei))\tPhi\cr
&+ (t-t^{1-k})t^{N_1-N_4+\Delta}(1-x^4)^\sfr \Phi'
+ (k+1) (t^{-k} |v_1| + |v_2|) \tPhi \cr
&+ \sum_{i\geq 1} {{k+1}\choose{i+1}} t^{i-k} \tPhi(L(i)v_1)
+ \sum_{i\geq 1} {{k+1}\choose{i+1}} \tPhi(L(i)v_2).\cr}$$
Note that
$$\frac{d\Phi}{dx} = \frac{d\Phi}{dt} \frac{dt}{dx} \qquad \hbox{and}\qquad
\frac{dt}{dx} = \frac{8x(1-x^2)}{(1+x^2)^3} = \frac{8x(1-x^4)}{(1+x^2)^4}$$
so that
$$\Phi' = \frac{d\Phi}{dt} = \frac{d\Phi}{dx} \frac{(1+x^2)^4}{8x(1-x^4)}.$$
{}From
$$\eqalign{ \frac{d}{dx} \tPhi(x) &= \frac{d}{dx}\left(
(1-x^4)^\sfr t^{N_1-N_4+\Delta} \Phi(t)\right) \cr
&= \frac{-x^3}{1-x^4}\tPhi(x)+\frac{N_1-N_4+\Delta}{t}\frac{dt}{dx}\tPhi(x)
+ (1-x^4)^\sfr t^{N_1-N_4+\Delta} \frac{d\Phi}{dx} \cr}$$
we get
$$(1-x^4)^\sfr t^{N_1-N_4+\Delta} \frac{d\Phi}{dt}
= \left(\frac{dt}{dx}\right)^{-1} \frac{d\tPhi}{dx} +
\frac{x^3}{1-x^4} \left(\frac{dt}{dx}\right)^{-1} \tPhi
- \frac{N_1-N_4+\Delta}{t} \tPhi.$$
This gives us
$$\eqalign{
&(t-t^{1-k}) t^{N_1-N_4+\Delta} (1-x^4)^\sfr \Phi'\cr
&= (1 - t^{-k}) \left(\frac{x(1+x^2)}{2(1-x^2)}
\frac{d}{dx} + \frac{x^4}{2(1-x^2)^2} + (N_4-N_1-\Delta)\right) \tPhi\cr
&= (1 - t^{-k}) (\cL_x + (N_4-N_1-\Delta)) \tPhi\cr}$$
which gives the result.$\hfill\bk$  \enddemo

Note that the operator $\cL_x$ and multiplication by any integral power of
$t^\shf$ preserve the ring $\cR_x$.

\pr{Lemma 19} For any $g\in {\Bbb C}(x)$, if $x = u^{-1}$, we have
$$\cL_x(xg(x^{-1})) = u^{-1} \cL_u(g(u)).$$ \epr

\demo{Proof} We have
$$\eqalign{
\cL_x(xg(x^{-1})) &= \frac{x(1+x^2)}{2(1-x^2)}
\left(x\frac{dg}{du}\frac{du}{dx}+g(x^{-1})
+ \frac{x^4}{1-x^4} g(x^{-1}) \right)\cr
&= \frac{u^2+1}{2u(u^2-1)} \left(-u\frac{dg}{du} + g(u)
+ \frac{1}{u^4-1} g(u)\right) \cr
&= \frac{1+u^2}{2(1-u^2)} \frac{dg}{du} + \frac{u^3}{2(1-u^2)^2} g(u) \cr
&= u^{-1} \cL_u(g(u)). \cr}$$
$\hfill\bk$  \enddemo

\pr{Lemma 20} For any $g\in {\Bbb C}(x)$, if $x = \frac{\pm y+\bi}{\bi y
\pm 1}$,
$t^{1/2} = \frac{2x}{1+x^2}$ and $N\in{\Bbb Z}$, we have
$$\cL_x(t^N (\bi y \pm 1)^{-1}g(x))
= \frac{-t^N}{\bi y \pm 1} (\cL_y -N-\sei)g\left(\frac{\pm y+\bi}{\bi y\pm
1}\right).$$
\epr

\demo{Proof} We have
$$\eqalign{
&\cL_x(t^N (\bi y \pm 1)^{-1}g(x))\cr
&=\frac{x(1+x^2)}{2(1-x^2)}\left[ \frac{t^N}{\bi y \pm 1} \frac{dg}{dy}
\frac{dy}{dx} + t^N g(x) \frac{d}{dx} (\bi y \pm 1)^{-1}\right. \cr
&\qquad\qquad
+N t^{N-1} \frac{dt}{dx} (\bi y \pm 1)^{-1} g(x)
+ t^N (\bi y \pm 1)^{-1}\left.  \frac{x^3}{1-x^4}
g(x)\right]\cr
&=\frac{x(1+x^2)}{2(1-x^2)}\left[ \frac{t^N}{\bi y \pm 1}
\frac{(\bi y \pm 1)^2}{2}
\frac{d}{dy} \ - t^N \frac{\bi}{2}\right. \cr
&\qquad\qquad
+N t^{N-1}  2\frac{2x}{1+x^2}\left. \frac{2(1-x^2)}{(1+x^2)^2}
(\bi y \pm 1)^{-1} + t^N (\bi y \pm 1)^{-1} \frac{x^3}{1-x^4}
\right] g(x) \cr
&=\frac{-t^N}{\bi y \pm 1}
\frac{y(1+y^2)}{2(1-y^2)}\left[ \frac{d}{dy} \ -\frac{\bi}{\bi y \pm 1}
-\frac{ 2N(1-y^2)}{y(1+y^2)}\right. \cr
&\qquad\qquad
+\left. \frac{(\pm y + \bi )^3}{\pm 4y \bi
(1-y^2) (\bi y \pm 1)}\right] g\left(\frac{\pm y+\bi}{\bi y \pm 1}\right)\cr
&=\frac{-t^N}{\bi y \pm 1}
\frac{y(1+y^2)}{2(1-y^2)} \left[ \frac{d}{dy} \
+ \frac{3y^2-1}{4y(1-y^2)}
-\frac{ 2N(1-y^2)}{y(1+y^2)}\right] g\left(\frac{\pm y+\bi}{\bi y \pm
1}\right)\cr}$$
$$\eqalign{
&=\frac{-t^N}{\bi y \pm 1}
\frac{y(1+y^2)}{2(1-y^2)} \left[ \frac{d}{dy} \ +\frac{y^3}{1-y^4}
- \frac{1-y^2}{4y(1+y^2)}
-\frac{ 2N(1-y^2)}{y(1+y^2)}\right] g\left(\frac{\pm y+\bi}{\bi y \pm
1}\right)\cr
&=\frac{-t^N}{\bi y \pm 1}
\frac{y(1+y^2)}{2(1-y^2)} \left[ \frac{d}{dy} \ +\frac{y^3}{1-y^4}
- \frac{(1-y^2)(1+8N)}{4y(1+y^2)}\right] g\left(\frac{\pm y+\bi}{\bi y \pm
1}\right)\cr
&=\frac{-t^N}{\bi y \pm 1}
\left[\cL_y - \frac{y(1+y^2)}{2(1-y^2)}
\frac{(1-y^2)(1+8N)}{4y(1+y^2)}\right] g\left(\frac{\pm y+\bi}{\bi y \pm
1}\right)\cr
&=\frac{-t^N}{\bi y \pm 1}
(\cL_y - N - \sei) g\left(\frac{\pm y+\bi}{\bi y \pm 1}\right)\cr}$$
$\hfill\bk$  \enddemo

\pr{Lemma 21} For any $g\in {\Bbb C}(x)$, if $x = \frac{w\pm 1}{\mp w+1}$,
$s^{1/2} = \frac{2w}{1-w^2}$ and $N\in{\Bbb Z}$, we have
$$\cL_x(s^{-N} (\mp w + 1)^{-1} g(x))
= \frac{-s^{-N}}{\mp w + 1} (s^{-1}\cL_w -(N+\sei)(1+s^{-1}))
g\left(\frac{w\pm 1}{\mp w + 1}\right).$$
\epr

\demo{Proof} We have
$$\eqalign{
&\cL_x(s^{-N} (\mp w + 1)^{-1} g(x)) \cr
&=\frac{x(1+x^2)}{2(1-x^2)}\bigg[ \frac{s^{-N}}{\mp w +1} \frac{dg}{dw}
\frac{dw}{dx} + s^{-N} g(x) \frac{dw}{dx}\frac{d}{dw}\frac{1}{\mp w+1} \cr
&\qquad\qquad
+\frac{1}{\mp w+1} g(x) \frac{dw}{dx} \frac{d}{dw} s^{-N}
+ \frac{s^{-N}}{(\mp w +1)} \frac{x^3}{(1-x^4)} g(x)\bigg]\cr
&=\frac{x(1+x^2)}{2(1-x^2)}\bigg[ \frac{s^{-N}}{\mp w +1}\frac{(\mp w +1)^2}{2}
\frac{d}{dw} \pm \frac{s^{-N}}{2} \cr
&\qquad\qquad
-\frac{1}{\mp w+1} \frac{(\mp w +1)^2}{2} Ns^{-N} \frac{2(1+w^2)}{w(1-w^2)}
+ \frac{s^{-N}}{(\mp w +1)} \frac{x^3}{(1-x^4)} \bigg] g(x)\cr
&=\frac{x(1+x^2)}{2(1-x^2)}\frac{s^{-N}}{\mp w +1}\bigg[\frac{(\mp w +1)^2}{2}
\frac{d}{dw} \pm \frac{\mp w +1}{2} \cr
&\qquad\qquad
- \frac{N(\mp w +1)^2(1+w^2)}{w(1-w^2)}
+ \frac{x^3}{1-x^4} \bigg] g(x)\cr
&=\frac{x(1+x^2)(\mp w +1)^2}{4(1-x^2)} \frac{s^{-N}}{\mp w +1}\bigg[
\frac{d}{dw} \pm \frac{1}{\mp w +1} \cr
&\qquad\qquad
- \frac{2N(1+w^2)}{w(1-w^2)}
+ \frac{2x^3}{(1-x^4)(\mp w +1)^2} \bigg] g(x)\cr
&= \frac{-s^{-N}}{\mp w +1} \frac{(1-w^2)(1+w^2)}{8w} \bigg[
\frac{d}{dw} \pm \frac{1}{\mp w +1} \cr
&\qquad\qquad
- \frac{2N(1+w^2)}{w(1-w^2)}
\mp \frac{(w\pm 1)^3}{4(\mp w +1)w(1+w^2)} \bigg] g(x)\cr}$$
$$\eqalign{
&= \frac{-s^{-N}}{\mp w +1} \frac{(1-w^2)(1+w^2)}{8w} \bigg[
\frac{d}{dw} + \frac{w^3}{1-w^4}
- \frac{(8N+1)(1+w^2)}{4w(1-w^2)} \bigg] g(x)\cr
&= \frac{-s^{-N}}{\mp w +1} \bigg[ \frac{(1-w^2)(1+w^2)}{8w}
\bigg(\frac{d}{dw} + \frac{w^3}{1-w^4}\bigg)
- (N+\sei) \frac{(1+w^2)^2}{4w^2} \bigg] g(x)\cr
&= \frac{-s^{-N}}{\mp w +1} \big[s^{-1} \cL_w
- (N+\sei) (1+s^{-1}) \big] g\left(\frac{w\pm 1}{\mp w+1}\right)\cr
}$$

$\hfill\bk$  \enddemo

\pr{Theorem 22} For any $k \in {\Bbb Z}$ we have
$$\eqalign{
&\tPsi(L(-k)v_4) = \cr
&\tPsi(L(k)v_3) - (N+\sei) s^{-k-1}[(1+s)^{k+1}-1]\tPsi
+ s^{-k-1} [(1+s)^k - 1] \cL_w \tPsi \cr
&+ \sum_{i\geq 0} {{k+1}\choose{i+1}} \left[s^{i-k} \tPsi(L(i)v_2)
+ \left(\frac{1+s}{s}\right)^{k-i} \tPsi(L(i)v_1)\right].\cr}$$
\epr

\demo{Proof} For any $k \in {\Bbb Z}$ we have
$$\eqalign{
&\tPsi(L(-k)v_4) \cr
&= (1-w^4)^\sfr z^{N-k+\sei} H(v_1,v_2,v_3,L(-k)v_4;z,z_2)\cr
&= (1-w^4)^\sfr z^{N-k+\sei} H(v_1,v_2,L(k)v_3,v_4;z,z_2) \cr
&+ (1-w^4)^\sfr z^{N-k+\sei} z_2^{k+1}\partial_2H(v_1,v_2,v_3,v_4;z,z_2)\cr
&+ (1-w^4)^\sfr z^{N-k+\sei} [(z_2 + z)^{k+1} - z_2^{k+1}]
\partial H(v_1,v_2,v_3,v_4;z,z_2)\cr
&+ (1-w^4)^\sfr z^{N-k+\sei} \sum_{i\geq 0} {{k+1}\choose{i+1}} z_2^{k-i}
H(v_1,L(i)v_2,v_3,v_4;z,z_2)\cr
&+ (1-w^4)^\sfr z^{N-k+\sei} \sum_{i\geq 0}{{k+1}\choose{i+1}}(z_2+z)^{k-i}
H(L(i)v_1,v_2,v_3,v_4;z,z_2) \cr
&= \tPsi(L(k)v_3) + (N_4-N_3-\bDel) s^{-k} \tPsi - s^{1-k}(1-w^4)^\sfr
s^{N_3-N_4+\bDel} \Psi' \cr
&- (N_1+N_2+\sei-\bDel) s^{-k-1}[(1+s)^{k+1} - 1] \tPsi \cr
&+ s^{-k} [(1+s)^{k+1} - 1] s^{N_3-N_4+\bDel} (1-w^4)^\sfr \Psi' \cr
&+ \sum_{i\geq 0} {{k+1}\choose{i+1}} s^{i-k} \tPsi(L(i)v_2)
+ \sum_{i\geq 0} {{k+1}\choose{i+1}} s^{i-k}(1+s)^{k-i}\tPsi(L(i)v_1)\cr}$$
Note that
$$\frac{d\Psi}{dw} = \frac{d\Psi}{ds} \frac{ds}{dw} \qquad \hbox{and}\qquad
\frac{ds}{dw} = \frac{8w(1+w^2)}{(1-w^2)^3} = \frac{8w(1-w^4)}{(1-w^2)^4}$$
so that
$$\Psi' = \frac{d\Psi}{ds} = \frac{d\Psi}{dw} \frac{(1-w^2)^4}{8w(1-w^4)}.$$
{}From
$$\eqalign{ \frac{d}{dw} \tPsi(w) &= \frac{d}{dw}\left(
(1-w^4)^\sfr s^{N_3-N_4+\bDel} \Psi(s)\right) \cr
&= \frac{-w^3}{1-w^4}\tPsi(w)+\frac{N_3-N_4+\bDel}{s}\frac{ds}{dw}\tPsi(w)
+ (1-w^4)^\sfr s^{N_3-N_4+\bDel} \frac{d\Psi}{dw} \cr}$$
we get
$$(1-w^4)^\sfr s^{N_3-N_4+\bDel} \frac{d\Psi}{ds}
= \left(\frac{ds}{dw}\right)^{-1} \frac{d\tPsi}{dw} +
\frac{w^3}{1-w^4} \left(\frac{ds}{dw}\right)^{-1} \tPsi
- \frac{N_3-N_4+\bDel}{s} \tPsi.$$
The two terms involving $\Psi'$ in the above expression for
$\tPsi(L(-k)v_4)$ combine to give
$$s^{-k}(1+s)[(1+s)^k - 1] (1-w^4)^\sfr s^{N_3-N_4+\bDel} \frac{d\Psi}{ds}$$
and since $1+s = \frac{(1+w^2)^2}{(1-w^2)^2}$ this gives us
$$\eqalign{
&s^{-k}(1+s)[(1+s)^k - 1] (1-w^4)^\sfr s^{N_3-N_4+\bDel} \Psi' \cr
&= s^{-k}[(1+s)^k - 1] \left(\frac{1-w^4}{8w} \frac{d}{dw} +
\frac{w^2}{8} - \frac{(N_3-N_4+\bDel)(1+s)}{s}\right) \tPsi\cr
&= s^{-k}[(1+s)^k - 1] \left(
\frac{\cL_w}{s} - \frac{(N_3-N_4+\bDel)(1+s)}{s}\right) \tPsi.\cr}$$
Substituting this in the earlier expression, after some simplification, we
get the result.$\hfill\bk$  \enddemo

Note that the operator $\cL_w$ and multiplication by any rational function
in $s^\shf$ preserve the ring $\cR_w$.

Now let us return to the problem of determining the general two-point
functions inductively.

\pr{Theorem 23} For $1\leq i \leq 4$ let $v_i \in \bV_{n_i}$ be eigenvectors
for $L(0)$ with $L(0)v_i = wt(v_i)v_i = |v_i| v_i$.
For $i = 1,2$ let $Y_i = Y(v_i,z_i)$, $\partial_i = \partial/\partial z_i$.
Then for any $k \in {\Bbb Z}$ we have
$$\eqalign{
&G(L(-k)v_1,v_2,v_3,v_4;z_1,z_2) \cr
&= \sum_{i\geq 0} {{k+i-2}\choose i} z_1^i G(v_1,v_2,v_3,L(k+i)v_4;z_1,z_2) \cr
&+ (-1)^k (z_1 - z_2)^{1-k} \partial_2 G(v_1,v_2,v_3,v_4;z_1,z_2)\cr
&+ (-1)^k \sum_{j\geq 0} (-1)^{j+1} {{1-k}\choose{j+1}} (z_1 - z_2)^{-k-j}
G(v_1,L(j)v_2,v_3,v_4;z_1,z_2) \cr
&+ (-1)^k \sum_{i\geq 0} {{k+i-1}\choose{i+1}}z_1^{-k-i}
G(v_1,v_2,L(i)v_3,v_4;z_1,z_2)\cr
&+ (-1)^k z_1^{1-k} (G(v_1,v_2,v_3,L(1)v_4;z_1,z_2) -
(\partial_1 + \partial_2) G(v_1,v_2,v_3,v_4;z_1,z_2)).\cr}$$
\epr

\demo{Proof} We have
$$\eqalign{
&G(L(-k)v_1,v_2,v_3,v_4;z_1,z_2) = (Y(L(-k)v_1,z_1)Y_2 v_3,v_4) \cr
&= \sum_{i\geq 0} {{k+i-2}\choose{i}} z_1^i (L(-k-i)Y_1 Y_2 v_3,v_4)\cr
&+ (-1)^k\sum_{i\geq 0}{{k+i-2}\choose{i}}z_1^{-k+1-i}(Y_1 L(i-1)Y_2
v_3,v_4)\cr
&= \sum_{i\geq 0} {{k+i-2}\choose{i}} z_1^i G(v_1,v_2,v_3,L(k+i)v_4;z_1,z_2)\cr
&+ (-1)^k \sum_{i\geq 0} {{k+i-2}\choose{i}} z_1^{-k+1-i}
\bigg[ (Y_1 [L(i-1),Y_2]v_3,v_4)\cr
&\qquad\qquad\qquad + (Y_1 Y_2 L(i-1) v_3,v_4) \bigg]\cr
&= \sum_{i\geq 0} {{k+i-2}\choose{i}} z_1^i G(v_1,v_2,v_3,L(k+i)v_4;z_1,z_2)\cr
&+ (-1)^k \sum_{i\geq 0} {{k+i-2}\choose{i}} z_1^{-k+1-i}
\sum_{j\geq 0} {i\choose j} z_2^{i-j} (Y_1 Y(L(j-1)v_2,z_2) v_3,v_4) \cr
&+ (-1)^k \sum_{i\geq 0} {{k+i-2}\choose{i}} z_1^{-k+1-i}
(Y_1 Y_2 L(i-1) v_3,v_4) \cr
&= \sum_{i\geq 0} {{k+i-2}\choose{i}} z_1^i G(L(k+i)v_4)
+ (-1)^k z_1^{1-k} \sum_{i\geq 0} {{k+i-2}\choose{i}}
\left(\frac{z_2}{z_1}\right)^i \partial_2 G \cr
&+ (-1)^k \sum_{j\geq 0} \sum_{i\geq 0} {{k+i-2}\choose{i}} {i\choose {j+1}}
z_1^{1-k} z_2^{-j-1} \left(\frac{z_2}{z_1}\right)^i G(L(j)v_2) \cr
&+ (-1)^k \sum_{i\geq 0} {{k+i-1}\choose{i+1}} z_1^{-k-i} G(L(i)v_3)
+ (-1)^k z_1^{1-k} (Y_1 Y_2 L(-1) v_3,v_4) \cr
&= \sum_{i\geq 0} {{k+i-2}\choose{i}} z_1^i G(L(k+i)v_4)
+ (-1)^k z_1^{1-k} \partial_2 G
\sum_{i\geq 0} {{k+i-2}\choose{i}} \left(\frac{z_2}{z_1}\right)^i \cr
&+ (-1)^k \sum_{j\geq 0} z_1^{-j-k} G(L(j)v_2)
\sum_{i\geq 0} {{k+i-2}\choose{i}} {i\choose {j+1}}
\left(\frac{z_2}{z_1}\right)^{i-j-1} \cr
&+ (-1)^k \sum_{i\geq 0} {{k+i-1}\choose{i+1}} z_1^{-k-i} G(L(i)v_3) \cr
&+ (-1)^k z_1^{1-k} (G(L(1)v_4) - ([L(-1),Y_1]Y_2 v_3,v_4) -
(Y_1[L(-1),Y_2] v_3,v_4)).\cr}$$
We may write the last line as
$$(-1)^k z_1^{1-k} (G(L(1)v_4) - (\partial_1 + \partial_2) G).$$
Also, note that
$${{k+i-2}\choose i} = (-1)^i \ {{1-k}\choose i}$$
so
$$\sum_{i\geq 0} {{k+i-2}\choose{i}} \left(\frac{z_2}{z_1}\right)^i =
\sum_{i\geq 0} (-1)^i {{1-k}\choose{i}} \left(\frac{z_2}{z_1}\right)^i =
\left(1 - \frac{z_2}{z_1}\right)^{1-k}$$
and therefore,
$$(-1)^k z_1^{1-k} \partial_2 G
\sum_{i\geq 0} {{k+i-2}\choose{i}} \left(\frac{z_2}{z_1}\right)^i
= (-1)^k (z_1 - z_2)^{1-k} \partial_2 G.$$
In addition, we have
$$\eqalign{
&\sum_{i\geq 0} (-1)^i {{1-k}\choose{i}} {i\choose{j+1}} t^{i-j-1} =
\frac{1}{(j+1)!} \partial_t^{j+1}
\sum_{i\geq 0} (-1)^i {{1-k}\choose{i}} t^i \cr
&= \frac{1}{(j+1)!} \partial_t^{j+1} (1 - t)^{1-k} =
(-1)^{j+1} {{1-k}\choose{j+1}} (1 - t)^{-j-k} \cr}$$
so that
$$\eqalign{
&(-1)^k \sum_{j\geq 0} z_1^{-j-k} G(L(j)v_2)
\sum_{i\geq 0} {{k+i-2}\choose{i}} {i\choose {j+1}}
\left(\frac{z_2}{z_1}\right)^{i-j-1} \cr
&= (-1)^k \sum_{j\geq 0} z_1^{-j-k} G(L(j)v_2)
\sum_{i\geq 0} (-1)^i {{1-k}\choose{i}} {i\choose {j+1}}
\left(\frac{z_2}{z_1}\right)^{i-j-1} \cr
&= (-1)^k \sum_{j\geq 0} z_1^{-j-k} G(L(j)v_2)
(-1)^{j+1} {{1-k}\choose{j+1}} \left(1 - \frac{z_2}{z_1}\right)^{-j-k} \cr
&= (-1)^k \sum_{j\geq 0} G(L(j)v_2)
(-1)^{j+1} {{1-k}\choose{j+1}} (z_1 - z_2)^{-j-k} \cr}$$
which gives the result.
$\hfill\bk$ \enddemo

\pr{Theorem 24} For $1\leq i \leq 4$ let $v_i \in \bV_{n_i}$ be eigenvectors
for $L(0)$ with $L(0)v_i = wt(v_i)v_i = |v_i| v_i$.
For $i = 1,2$ let $Y_i = Y(v_i,z_i)$, $\partial_i = \partial/\partial z_i$.
Then for any $k \in {\Bbb Z}$ we have
$$\eqalign{
&G(v_1,L(-k)v_2,v_3,v_4;z_1,z_2)
= \sum_{i\geq 0} {{k+i-2}\choose i} z_2^i G(v_1,v_2,v_3,L(k+i)v_4;z_1,z_2)\cr
&+ (-1)^k \sum_{i\geq 0} {{k+i-1}\choose{i+1}}z_2^{-k-i}
G(v_1,v_2,L(i)v_3,v_4;z_1,z_2)
-(\partial_1 G) (z_1 - z_2)^{1-k} \cr
&+ \sum_{j\geq 0}G(L(j)v_1,v_2,v_3,v_4;z_1,z_2) (-1)^j z_2^{-j-k}\cdot\cr
&\qquad\qquad\qquad
\cdot\sum_{i\geq 0} (-1)^i {{1-k}\choose i} {{i+j+k-1}\choose{j+1}}
\left(\frac{z_2}{z_1}\right)^{i+j+k} \cr
&+ (-1)^k z_2^{1-k} (G(v_1,v_2,v_3,L(1)v_4;z_1,z_2) -
(\partial_1 + \partial_2) G(v_1,v_2,v_3,v_4;z_1,z_2)).\cr}$$
\epr

\demo{Proof} We have
$$\eqalign{
G(&v_1,L(-k)v_2,v_3,v_4;z_1,z_2) = (Y_1Y(L(-k)v_2,z_2) v_3,v_4) \cr
&= \sum_{i\geq 0} {{k+i-2}\choose{i}} z_2^i (Y_1 L(-k-i) Y_2 v_3,v_4)\cr
&+ (-1)^k\sum_{i\geq 0}{{k+i-2}\choose{i}}z_2^{-k+1-i}(Y_1Y_2 L(i-1)
v_3,v_4)\cr
&= \sum_{i\geq 0} {{k+i-2}\choose{i}} z_2^i (L(-k-i)Y_1 Y_2 v_3,v_4)\cr
&- \sum_{i\geq 0} {{k+i-2}\choose{i}} z_2^i ([L(-k-i),Y_1]Y_2 v_3,v_4)\cr
&+ (-1)^k \sum_{i\geq 0} {{k+i-1}\choose{i+1}} z_2^{-k-i} G(L(i)v_3) \cr
&+ (-1)^k z_2^{1-k} (G(L(1)v_4) - (\partial_1 + \partial_2) G)\cr
&= \sum_{i\geq 0} {{k+i-2}\choose{i}} z_2^i G(L(k+i)v_4)\cr
&- \sum_{i\geq 0} {{k+i-2}\choose{i}} z_2^{i}
\sum_{j\geq 0} {{-k-i+1}\choose j} z_1^{1-i-j-k}(Y(L(j-1)v_1,z_1)Y_2
v_3,v_4)\cr
&+ (-1)^k \sum_{i\geq 0} {{k+i-1}\choose{i+1}} z_2^{-k-i} G(L(i)v_3) \cr
&+ (-1)^k z_2^{1-k} (G(L(1)v_4) - (\partial_1 + \partial_2) G).\cr}$$
We may rewrite the second line in the last expression as
$$\eqalign{
&- \sum_{j\geq 0}G(L(j-1)v_1) (-1)^j z_2^{1-j-k}\cdot\cr
&\qquad\qquad\qquad\cdot\sum_{i\geq 0} (-1)^i
{{1-k}\choose{i}} {{i+j+k-2}\choose j}\left(\frac{z_2}{z_1}\right)^{i+j+k-1}\cr
&= -(\partial_1 G) (z_1 - z_2)^{1-k} \cr
&+ \sum_{j\geq 0}G(L(j)v_1) (-1)^j z_2^{-j-k} \sum_{i\geq 0} (-1)^i
{{1-k}\choose{i}} {{i+j+k-1}\choose{j+1}}\left(\frac{z_2}{z_1}\right)^{i+j+k}
\cr}$$
giving the result. Note that the $j = 0$ term in the last line simplifies to
\newline $(k - 1) |v_1| G (z_1 - z_2)^{-k}$.
$\hfill\bk$  \enddemo

\pr{Theorem 25} For any $k\in {\Bbb Z}$ we have
$$\eqalign{
&H(L(-k)v_1) = \cr
&\sum_{i\geq 0}(-1)^i {{1-k}\choose i}(z_2+z)^i H(L(k+i)v_4)\cr
&+ (-1)^k (z_2 + z)^{1-k} (H(L(1)v_4) - \partial_2 H) \cr
&+ (-1)^k \sum_{j\geq 0} H(L(j)v_3) z^{-j-k} \sum_{i\geq 0} {{1-k}\choose i}
{{i+j+k-1}\choose{j+1}} \left(\frac{z}{z_2}\right)^{i+j+k}\cr
&+ (-1)^k \sum_{i\geq 0} {{k+i-1}\choose{i+1}} z^{-k-i} H(L(i)v_2) \cr
&+ (-1)^k z^{1-k} (\partial_2 H - \partial H) \cr}$$
and
$$\eqalign{
H(L(-k)v_2) &= \sum_{i\geq 0} {{k+i-2}\choose i} z_2^i H(L(k+i)v_4)\cr
&+ (-1)^k z_2^{1-k} (H(L(1)v_4) - \partial_2 H) \cr
&+ (-1)^k \sum_{i\geq 0} {{k+i-1}\choose{i+1}} z_2^{-k-i} H(L(i)v_3) \cr
&- z^{1-k} \partial H - \sum_{i\geq 0} {{1-k}\choose{i+1}} z^{-k-i}
H(L(i)v_1) \cr}$$
\epr

\pr{Theorem 26} For any $k \in {\Bbb Z}$ we have
$$\eqalign{
\tPhi(L(-k)v_1) &= \sum_{i\geq 0}{{k+i-2}\choose i}t^{-i}\tPhi(L(k+i)v_4)\cr
&+ (-1)^k (N+\sei) t^{k-1} [1 - (1-t)^{1-k}] \tPhi \cr
&+ (-1)^k t^{k-1} (1-t) [(1-t)^{-k} - 1] \cL_x \tPhi \cr
&+ (-1)^k \sum_{j\geq 0} (-1)^{j+1} {{1-k}\choose{j+1}}
\left(\frac{t}{1-t}\right)^{j+k} \tPhi(L(j)v_2)\cr
&+ (-1)^k \sum_{i\geq 0}{{k+i-1}\choose{i+1}} t^{k+i} \tPhi(L(i)v_3)\cr
&+ (-1)^k t^{k-1} \tPhi(L(1)v_4)\cr}$$
and
$$\eqalign{
\tPhi(L(-k)v_2) &= \sum_{i\geq 0}{{k+i-2}\choose i} \tPhi(L(k+i)v_4)\cr
&+ (-1)^k \sum_{i\geq 0} {{k+i-1}\choose{i+1}} \tPhi(L(i)v_3)\cr
&+ [t^k (1-t)^{-k} - (-1)^k] (1-t) \cL_x \tPhi \cr
&+ \sum_{j\geq 0} (-1)^j \tPhi(L(j)v_1) \sum_{i\geq 0} (-1)^i
{{1-k}\choose i} {{i+j+k-1}\choose{j+1}} t^{i+j+k} \cr
&+ (-1)^k \tPhi(L(1)v_4) + (-1)^k (N+\sei) \tPhi.\cr}$$
\epr

\pr{Theorem 27} For any $k\in {\Bbb Z}$ we have
$$\eqalign{
\tPsi(L(-k)v_1) &= \sum_{i\geq 0}(-1)^i {{1-k}\choose i}
\left(\frac{1+s}{s}\right)^i \tPsi(L(k+i)v_4)\cr
&+ (-1)^k \left(\frac{1+s}{s}\right)^{1-k} \tPsi(L(1)v_4) \cr
&+ (-1)^k \sum_{j\geq 0} \tPsi(L(j)v_3) \sum_{i\geq 0} {{1-k}\choose i}
{{i+j+k-1}\choose{j+1}} s^{i+j+k}\cr
&+ (-1)^k \left[\left(\frac{s}{1+s}\right)^k - 1\right] \cL_w \tPsi \cr
&+ (-1)^k \sum_{i\geq 0} {{k+i-1}\choose{i+1}} \tPsi(L(i)v_2)
+ (-1)^k (N+\sei) \tPsi\cr}$$
and
$$\eqalign{
\tPsi(L(-k)v_2) &= \sum_{i\geq 0} {{k+i-2}\choose i} s^{-i}\tPsi(L(k+i)v_4)\cr
&+ (-1)^k s^{k-1} \tPsi(L(1)v_4) + (N+\sei) \tPsi \cr
&+ (-1)^k \sum_{i\geq 0} {{k+i-1}\choose{i+1}} s^{k+i} \tPsi(L(i)v_3) \cr
&- \sum_{i\geq 0} {{1-k}\choose{i+1}} \tPsi(L(i)v_1)
+ \frac{1}{1+s}[(-1)^k s^k -1] \cL_w \tPsi.}$$
\epr

\pr{Theorem 28} (Rationality) For $1\leq i \leq 4$ let $v_i \in \bV_{n_i}$ be
eigenvectors for $L(0)$ with $L(0)v_i = wt(v_i)v_i = |v_i| v_i$. Suppose
that $n_i \in\{1,3\}$, $n_4 = n_1+n_2+n_3$ (mod 4), $wt(v_i) = N_i +
\Delta_{n_i}$ and $N = N_1+N_2+N_3-N_4$. Then, after the substitution
$(z_2/z_1)^{1/2} = 2x/(1+x^2)$, the series
$${\tPhi}(v_1,v_2,v_3,v_4;x) = (1-x^4)^\sfr z_2^{N+\sei}
G(v_1,v_2,v_3,v_4;z_1,z_2)$$
converges absolutely in the domain $|x| < \sqrt{3-2\sqrt{2}}$ to a rational
function in the ring $\cR_x = {\Bbb C}[x,x^{-1}(x^4-1)^{-1}]$. \epr

\demo{Proof} We proof the statement by induction on $M = N_1+N_2+N_3+N_4$.
The base case, when $N_i = 0$ for $1\leq i\leq 4$, is given by Theorem 13.
The result of Theorem 18 with $k > 0$ inductively reduces the general case
to the case where $N_4 = 0$, and with $k < 0$ it reduces further to the case
where $N_3 = 0$ also. The two results of Theorem 26 with $k > 0$ further
reduce one to the base case.$\hfill\bk$  \enddemo

\pr{Theorem 29} For any $v_1,v_2,v_3\in \bV_1$, $v_4\in\bV_3$, we have
$$[\tPhi]_0 \sim \frac{1}{x_\infty} B_\infty [\tPhi]_\infty.$$ \epr

\demo{Proof} It sufficies to prove this for homogeneous vectors $v_i$ with
$wt(v_i) = N_i + \Delta_{n_i}$, where $v_i\in\bV_{n_i}$.
We will prove this using Theorem 18, Theorem 26, and Lemma 19,
by induction on $N_1+N_2+N_3+N_4$. The base case, when all $N_i = 0$, has
been established already.
First note that since $wt(v_i) = wt(\theta(v_i))$, the $N$ in Theorems
18 and 26 is the same for each of the entries in the matrices $[\tPhi]_0$
and $[\tPhi]_\infty$.
Also, $t = 4x_0^2/(1+x_0^2)^2 = 4x_\infty^2/(1+x_\infty^2)^2$, so the
$t$ in Theorems 18 and 26 is the same whether $x$ is $x_0$ or $x_\infty$.
First we will do the inductive step which reduces the weight of $v_4$.
Assume the statement is true as stated for all
choices of four vectors whose weights add up to be less than or equal to
$N_1+N_2+N_3+N_4$. We will show that it is then true with $v_4$ replaced by
$L(-k)v_4$ for any $0 < k \in {\Bbb Z}$. We have
$$\eqalign{
&[\tPhi(L(-k)v_4)]_0 = \cr
&[\tPhi(L(k)v_3)]_0 - (N+\sei)[\tPhi]_0 + (1 - t^{-k})
\cL_{x_0} [\tPhi]_0 \cr
&+ \sum_{i\geq 0} {{k+1}\choose{i+1}} \left(t^{i-k} [\tPhi(L(i)v_1)]_0
+ [\tPhi(L(i)v_2)]_0 \right) \cr
&\sim \frac{1}{x_\infty} B_\infty [\tPhi(L(k)v_3)]_\infty -
(N+\sei) \frac{1}{x_\infty} B_\infty [\tPhi]_\infty + (1 - t^{-k})
\cL_{x_0} \frac{1}{x_\infty} B_\infty [\tPhi]_\infty \cr
&+ \sum_{i\geq 0} {{k+1}\choose{i+1}}
\frac{1}{x_\infty} B_\infty \left(t^{i-k} [\tPhi(L(i)v_1)]_\infty
+ [\tPhi(L(i)v_2)]_\infty \right) \cr
&= \frac{1}{x_\infty} B_\infty \left[ [\tPhi(L(k)v_3)]_\infty -
(N+\sei) [\tPhi]_\infty + (1 - t^{-k})
\cL_{x_\infty} [\tPhi]_\infty\right] \cr
&+\frac{1}{x_\infty} B_\infty \sum_{i\geq 0} {{k+1}\choose{i+1}}
\left(t^{i-k} [\tPhi(L(i)v_1)]_\infty
+ [\tPhi(L(i)v_2)]_\infty \right) \cr
&= \frac{1}{x_\infty} B_\infty [\tPhi(L(-k)v_4)]_\infty. \cr}$$
But the same calculation with $0 > k \in {\Bbb Z}$ shows that the statement
is also true with $v_3$ replaced by $L(k)v_3$.

Applying the two parts of Theorem 26 for $0 < k \in {\Bbb Z}$, one similarly
gets that the statement is true with $v_i$ replaced by $L(-k)v_i$, for $i =
1,2$. Lemma 19 plays a crucial role, giving
$$\cL_{x_0} \frac{1}{x_\infty} [\tPhi]_\infty
= \frac{1}{x_\infty} \cL_{x_\infty} [\tPhi]_\infty .$$
$\hfill\bk$ \enddemo

\pr{Theorem 30} Let $v_1,v_2,v_3\in \bV_1$ and $v_4\in\bV_3$ with
$wt(v_i) = |v_i| = N_i + \Delta_{n_i}$ and $N = N_1 + N_2 + N_3 - N_4$.
Then we have
$$[\tPhi]_0 \sim \frac{t^N}{\bi x_\bi + 1} B_\bi [\tPhi]_\bi
\sim \frac{t^N}{\bi x_{-\bi} - 1} B_{-\bi} [\tPhi]_{-\bi}.$$ \epr

\demo{Proof}
We will prove this using Theorem 18, Theorem 26, and Lemma 20,
by induction on $N_1+N_2+N_3+N_4$. The base case, when all $N_i = 0$, has
been established already.
Since $wt(v_i) = wt(\theta(v_i))$, the $N$ in Theorems
18 and 26 is the same for each of the entries in the matrices $[\tPhi]_0$
and $[\tPhi]_{\pm\bi}$.
First we will do the inductive step which reduces the weight of $v_4$.
Assume the statement is true as stated for all
choices of four vectors whose weights add up to be less than or equal to
$N_1+N_2+N_3+N_4$. We will show that it is then true with $v_4$ replaced by
$L(-k)v_4$ for any $0 < k \in {\Bbb Z}$. We have
$$\eqalign{
&[\tPhi(L(-k)v_4)]_0\cr
&= [\tPhi(L(k)v_3)]_0 - (N+\sei)[\tPhi]_0 + (1 - t^{-k})
\cL_{x_0} [\tPhi]_0 \cr
&+ \sum_{i\geq 0} {{k+1}\choose{i+1}} \left(t^{i-k} [\tPhi(L(i)v_1)]_0
+ [\tPhi(L(i)v_2)]_0 \right) \cr
&\sim \frac{t^{N-k}}{\bi x_\bi +1} B_\bi [\tPhi(L(k)v_3)]_\bi -
(N+\sei) \frac{t^{N}}{\bi x_\bi +1} B_\bi [\tPhi]_\bi + (1 - t^{-k})
\cL_{x_0} \frac{t^{N}}{\bi x_\bi +1} B_\bi [\tPhi]_\bi \cr
&+ \sum_{i\geq 0} {{k+1}\choose{i+1}}
\frac{t^{N-i}}{\bi x_\bi +1} B_\bi \left(t^{i-k} [\tPhi(L(i)v_1)]_\bi
+ [\tPhi(L(i)v_2)]_\bi \right) \cr
&= \frac{t^{N-k}}{\bi x_\bi +1} B_\bi \bigg[ [\tPhi(L(k)v_3)]_\bi -
(N\!+\sei) t^k [\tPhi]_\bi + (1 - t^{-k})
\left(-t^k (\cL_{x_\bi}\! -N\!-\sei) [\tPhi]_\bi\right) \cr
&+ \sum_{i\geq 0} {{k+1}\choose{i+1}}
\left([\tPhi(L(i)v_1)]_\bi
+ t^{k-i} [\tPhi(L(i)v_2)]_\bi \right)\bigg] \cr
&= \frac{t^{N-k}}{\bi x_\bi +1} B_\bi \bigg[ [\tPhi(L(k)v_3)]_\bi
+ (1 - t^k) \cL_{x_\bi} [\tPhi]_\bi - (N+\sei) [\tPhi]_\bi \cr
&+ \sum_{i\geq 0} {{k+1}\choose{i+1}}
\left([\tPhi(L(i)v_1)]_\bi
+ t^{k-i} [\tPhi(L(i)v_2)]_\bi \right)\bigg] \cr
&= \frac{t^{N-k}}{\bi x_\bi +1} B_\bi [\tPhi(L(-k)v_4)]_\bi.\cr}$$

In the last step we used Theorem 18 with $v_1$ and $v_2$ switched, $z_1$ and
$z_2$ switched, and therefore, $t$ replaced by $t^{-1}$.
The same calculation with $0 > k \in {\Bbb Z}$ shows that the statement
is also true with $v_3$ replaced by $L(k)v_3$.

We now apply the two parts of Theorem 26 for $0<k\in{\Bbb Z}$ to
get that the statement is true with $v_i$ replaced by $L(-k)v_i$, for $i =
1,2$. Lemma 20 plays a crucial role. We have
$$\eqalign{
&[\tPhi(L(-k)v_1)]_0 \cr
&= \sum_{i\geq 0}{{k+i-2}\choose i}t^{-i}[\tPhi(L(k+i)v_4)]_0\cr
&+ (-1)^k (N+\sei) t^{k-1} [1 - (1-t)^{1-k}] [\tPhi]_0 \cr
&+ (-1)^k t^{k-1} (1-t) [(1-t)^{-k} - 1] \cL_{x_0} [\tPhi]_0 \cr
&+ (-1)^k \sum_{j\geq 0} (-1)^{j+1} {{1-k}\choose{j+1}}
\left(\frac{t}{1-t}\right)^{j+k} [\tPhi(L(j)v_2)]_0\cr
&+ (-1)^k \sum_{i\geq 0}{{k+i-1}\choose{i+1}} t^{k+i} [\tPhi(L(i)v_3)]_0
+ (-1)^k t^{k-1} [\tPhi(L(1)v_4)]_0\cr}$$
$$\eqalign{
&\sim \sum_{i\geq 0}{{k+i-2}\choose i}t^{-i} \frac{t^{N+k+i}}{\bi x_\bi +1}
B_\bi [\tPhi(L(k+i)v_4)]_\bi\cr
&+ (-1)^k (N+\sei) t^{k-1} [1 - (1-t)^{1-k}] \frac{t^N}{\bi x_\bi +1}
B_\bi [\tPhi]_\bi \cr
&+ (-1)^k t^{k-1} (1-t) [(1-t)^{-k} - 1] \cL_{x_0} \frac{t^N}{\bi x_\bi +1}
B_\bi [\tPhi]_\bi \cr
&+ (-1)^k \sum_{j\geq 0} (-1)^{j+1} {{1-k}\choose{j+1}}
\left(\frac{t}{1-t}\right)^{j+k} \frac{t^{N-j}}{\bi x_\bi +1}
B_\bi [\tPhi(L(j)v_2)]_\bi\cr
&+ (-1)^k \sum_{i\geq 0}{{k+i-1}\choose{i+1}} t^{k+i}
\frac{t^{N-i}}{\bi x_\bi +1} B_\bi [\tPhi(L(i)v_3)]_\bi\cr
&+ (-1)^k t^{k-1} \frac{t^{N+1}}{\bi x_\bi +1} B_\bi [\tPhi(L(1)v_4)]_\bi\cr
&= \sum_{i\geq 0}{{k+i-2}\choose i} \frac{t^{N+k}}{\bi x_\bi +1}
B_\bi [\tPhi(L(k+i)v_4)]_\bi\cr
&+ (-1)^k (N+\sei) t^{k-1} [1 - (1-t)^{1-k}] \frac{t^N}{\bi x_\bi +1}
B_\bi [\tPhi]_\bi \cr
&+ (-1)^k t^{k-1} (1-t) [(1-t)^{-k} - 1] \left[\frac{-t^N}{\bi x_\bi +1}
B_\bi (\cL_{x_\bi} - N - \sei)\right]  [\tPhi]_\bi \cr
&+ (-1)^k \sum_{j\geq 0} (-1)^{j+1} {{1-k}\choose{j+1}}
(1-t)^{-j-k} \frac{t^{N+k}}{\bi x_\bi +1}
B_\bi [\tPhi(L(j)v_2)]_\bi\cr
&+ (-1)^k \sum_{i\geq 0}{{k+i-1}\choose{i+1}}
\frac{t^{N+k}}{\bi x_\bi +1} B_\bi [\tPhi(L(i)v_3)]_\bi
+ (-1)^k \frac{t^{N+k}}{\bi x_\bi +1} B_\bi [\tPhi(L(1)v_4)]_\bi\cr
&= \frac{t^{N+k}}{\bi x_\bi +1} B_\bi\bigg[ \sum_{i\geq 0}{{k+i-2}\choose i}
[\tPhi(L(k+i)v_4)]_\bi\cr
&+ (-1)^k (N+\sei) t^{-1} [1 - (1-t)^{1-k}]  [\tPhi]_\bi \cr
&- (-1)^k t^{-1} (1-t) [(1-t)^{-k} - 1]
(\cL_{x_\bi} - N - \sei) [\tPhi]_\bi \cr
&+ (-1)^k \sum_{j\geq 0} (-1)^{j+1} {{1-k}\choose{j+1}}
(1-t)^{-j-k} [\tPhi(L(j)v_2)]_\bi\cr
&+ (-1)^k \sum_{i\geq 0}{{k+i-1}\choose{i+1}} [\tPhi(L(i)v_3)]_\bi
+ (-1)^k [\tPhi(L(1)v_4)]_\bi\bigg] \cr
&= \frac{t^{N+k}}{\bi x_\bi +1} B_\bi\bigg[ \sum_{i\geq 0}{{k+i-2}\choose i}
[\tPhi(L(k+i)v_4)]_\bi\cr
&+ (-1)^k (N+\sei) [\tPhi]_\bi
+ [t^{-k} (1-t^{-1})^{-k} - (-1)^k] (1 - t^{-1})
\cL_{x_\bi} [\tPhi]_\bi \cr
&+ \sum_{j\geq 0} (-1)^j [\tPhi(L(j)v_2)]_\bi \sum_{i\geq 0} {{1-k}\choose{i}}
{{i+j+k-1}\choose{j+1}} t^{-i-j-k} \cr}$$
$$\eqalign{
&+ (-1)^k \sum_{i\geq 0}{{k+i-1}\choose{i+1}} [\tPhi(L(i)v_3)]_\bi
+ (-1)^k [\tPhi(L(1)v_4)]_\bi\bigg] \cr
&= \frac{t^{N+k}}{\bi x_\bi +1} B_\bi [\tPhi(L(-k)v_1)]_\bi.\cr}$$
In the last step we used the second part of Theorem 26 with $v_1$ and $v_2$
switched, $z_1$ and $z_2$ switched, and therefore, $t$ replaced by $t^{-1}$.

The calculation for $[\tPhi(L(-k)v_2)]_0$ is similar, completing the
inductive proof of the first part of the theorem. The second part, with
$x_{-\bi}$ in place of $x_\bi$, $B_{-\bi}$ in place of $B_\bi$, and certain
sign changes, is proved in exactly the same way. $\hfill\bk$ \enddemo

\pr{Theorem 31} Let $v_1,v_2,v_3\in \bV_1$ and $v_4\in\bV_3$ with
$wt(v_i) = |v_i| = N_i + \Delta_{n_i}$ and $N = N_1 + N_2 + N_3 - N_4$.
Then we have
$$[\tPhi]_0 \sim \frac{s^{-N}}{-x_1 + 1} B_1 [\tPsi]_1
\sim \frac{s^{-N}}{x_{-1} + 1} B_{-1} [\tPsi]_{-1}.$$ \epr

\demo{Proof}
We will prove this using Theorems 18, 22, 27, and Lemma 21,
by induction on $N_1+N_2+N_3+N_4$. The base case, when all $N_i = 0$, has
been established already.
First note that since $wt(v_i) = wt(\theta(v_i))$, the $N$ in Theorems
18 and 27 is the same for each of the entries in the matrices $[\tPhi]_0$
and $[\tPsi]_{\pm 1}$.
First we will do the inductive step which reduces the weight of $v_4$.
Assume the statement is true as stated for all
choices of four vectors whose weights add up to be less than or equal to
$N_1+N_2+N_3+N_4$. We will show that it is then true with $v_4$ replaced by
$L(-k)v_4$ for any $0 < k \in {\Bbb Z}$. We have
$$\eqalign{
&[\tPhi(L(-k)v_4)]_0 \cr
&= [\tPhi(L(k)v_3)]_0 - (N+\sei)[\tPhi]_0 + (1 - t^{-k})
\cL_{x_0} [\tPhi]_0 \cr
&+ \sum_{i\geq 0} {{k+1}\choose{i+1}} \left(t^{i-k} [\tPhi(L(i)v_1)]_0
+ [\tPhi(L(i)v_2)]_0 \right) \cr
&\sim \frac{s^{-N+k}}{-x_1+1} B_1 [\tPsi(L(k)v_3)]_1 -
(N+\sei) \frac{s^{-N}}{-x_1+1} B_1 [\tPsi]_1\cr
&+ (1 - t^{-k}) \cL_{x_0} \frac{s^{-N}}{-x_1+1} B_1 [\tPsi]_1 \cr
&+ \sum_{i\geq 0} {{k+1}\choose{i+1}}
\frac{s^{-N+i}}{-x_1+1} B_1 \left(t^{i-k} [\tPsi(L(i)v_1)]_1
+ [\tPsi(L(i)v_2)]_1 \right) \cr
&= \frac{s^{-N+k}}{-x_1+1} B_1 \bigg[ [\tPsi(L(k)v_3)]_1 -
(N+\sei) s^{-k} [\tPsi]_1 \cr
&- (1 - t^{-k}) s^{-k}
\left(s^{-1}\cL_{x_1}-(N+\sei)(1+s^{-1})\right) [\tPsi]_1 \cr
&+ \sum_{i\geq 0} {{k+1}\choose{i+1}}
\left[\left(\frac{s+1}{s}\right)^{k-i} [\tPsi(L(i)v_1)]_1
+ s^{i-k} [\tPsi(L(i)v_2)]_1 \right] \bigg] \cr}$$
$$\eqalign{
&= \frac{s^{-N+k}}{-x_1+1} B_1 \bigg[ [\tPsi(L(k)v_3)]_1 -
(N+\sei) s^{-k} [\tPsi]_1
+ ((1+s)^k - 1) s^{-k-1} \cL_{x_1} [\tPsi]_1 \cr
&+ (1 - (1+s)^k) s^{-k} (1+s^{-1}) (N+\sei) [\tPsi]_1 \cr
&+ \sum_{i\geq 0} {{k+1}\choose{i+1}}
\left[\left(\frac{s+1}{s}\right)^{k-i} [\tPsi(L(i)v_1)]_1
+ s^{i-k} [\tPsi(L(i)v_2)]_1 \right] \bigg] \cr
&= \frac{s^{-N+k}}{-x_1+1} B_1 \bigg[ [\tPsi(L(k)v_3)]_1
- (N+\sei) s^{-k-1} ((1+s)^{k+1} - 1) [\tPsi]_1 \cr
&+ s^{-k-1} ((1+s)^k - 1) \cL_{x_1} [\tPsi]_1 \cr
&+ \sum_{i\geq 0} {{k+1}\choose{i+1}}
\left[\left(\frac{s+1}{s}\right)^{k-i} [\tPsi(L(i)v_1)]_1
+ s^{i-k} [\tPsi(L(i)v_2)]_1 \right] \bigg] \cr
&= \frac{s^{-N+k}}{-x_1+1} B_1 [\tPsi(L(-k)v_4)]_1 \cr}$$
We used Theorem 22 in the last step.

The same calculation with $0 > k \in {\Bbb Z}$ shows that the statement
is also true with $v_3$ replaced by $L(k)v_3$.

Applying the two parts of Theorem 27 for $0 < k \in {\Bbb Z}$, one similarly
gets that the statement is true with $v_i$ replaced by $L(-k)v_i$, for $i =
1,2$. Lemma 21 plays a crucial role.
$\hfill\bk$ \enddemo

We are, at last, ready to state the new ``matrix'' Jacobi-Cauchy Identity which
is valid for the $c = \shf$ minimal model. This is the main objective of this
paper, but considerable further work remains to be done in order to understand
other minimal models and the WZW models. That will be the subject of future
investigations.

\pr{Theorem 32} (Matrix Jacobi-Cauchy Identity)
Let $v_1,v_2,v_3\in \bV_1$ and $v_4\in\bV_3$ with
$wt(v_i) = |v_i| = N_i + \Delta_{n_i}$ and $N = N_1 + N_2 + N_3 - N_4$.
Let $f(x)$ be any function in $\cR_x$. Let $C_0$ be a small positively oriented
circle with center $x_0 = 0$ and for $\alpha\in\{\infty,1,-1,\bi,-\bi\}$ let
$C_\alpha$ be the circle obtained from $C_0$ by the appropriate M\"obius
transformation sending $0$ to $\alpha$.
Then we have
$$\eqalign{
0 = &\oint_{C_0} [\Phi]_0 f(x_0) d x_0 \cr
&+ \oint_{C_\infty} \frac{-1}{x_\infty^3} B_\infty [\tPhi]_\infty
f(1/x_\infty)\ d x_\infty \cr
&+ \oint_{C_\bi} \frac{2t^N}{(\bi x_\bi + 1)^3} B_\bi [\tPhi]_\bi
f\left(\frac{x_\bi + \bi}{\bi x_\bi + 1}\right) \ d x_\bi \cr
&+ \oint_{C_{-\bi}} \frac{2t^N}{(\bi x_{-\bi} - 1)^3} B_{-\bi} [\tPhi]_{-\bi}
f\left(\frac{-x_{-\bi} + \bi}{\bi x_{-\bi} - 1}\right) \ d x_{-\bi} \cr
&+ \oint_{C_1} \frac{2s^{-N}}{(-x_1 + 1)^3} B_1 [\tPsi]_1
f\left(\frac{x_1 + 1}{-x_1 + 1}\right) \ d x_1 \cr
&+ \oint_{C_{-1}} \frac{2s^{-N}}{(x_{-1} + 1)^3} B_{-1} [\tPsi]_{-1}
f\left(\frac{x_{-1} - 1}{x_{-1} + 1}\right) \ d x_{-1}. \cr}$$
\epr

\demo{Proof} Apply the Cauchy residue theorem to $f(x)$ times the globally
defined matrix valued function $G(x)$ to which each of the expressions in
Theorems 29 - 31 converge. It says that the sum of the residues at the six
possible poles is zero, that is,
$$0 = \sum_\alpha \oint_{C_\alpha} G(x_0)f(x_0)dx_0.$$
Let
$$x_0 = \mu_\alpha(x_\alpha)
= \frac{a_\alpha x_\alpha + b_\alpha}{c_\alpha x_\alpha + d_\alpha}$$
denote the M\"obius transformation chosen before Theorem 13 to relate $x_0$ to
$x_\alpha$. Then the chain rule gives
$dx_0 = \frac{a_\alpha d_\alpha - b_\alpha c_\alpha}{(c_\alpha x_\alpha +
d_\alpha)^2}\ dx_\alpha$. For each $\alpha$ we write the corresponding
integral with $G(x_0)f(x_0)$ represented by the series in $x_\alpha$
which converges to it in the neighborhood of $x_0 = \alpha$, and we write
$dx_0$ as the appropriate chain rule factor times $dx_\alpha$.
$\hfill\bk$
\enddemo

Just as Corollary 3 was obtained from Theorem 2, we can
obtain infinitely many explicit identities for products
of components of intertwining operators from Theorem 32 by making
explicit choices for the test function $f(x)$. One computes the integral
in each term of the Matrix Jacobi-Cauchy Identity (MJCI) by expanding the
integrand as a series in the local variable $x_\alpha$ and then finding the
residue as the coefficient of $x_\alpha^{-1}$. The algebra involved is tedious
but straightforward, so we will not present it here. The most general function
$f(x)\in\cR_x$ is a linear combination of functions of the form
$$f(x) = x^{m_0} (1+\bi x)^{m_\bi} (1-\bi x)^{m_{-\bi}} (1-x)^{m_1}
(1+x)^{m_{-1}}$$
where $m_0, m_\bi, m_{-\bi}, m_1, m_{-1} \in {\Bbb Z}$ are five independent
parameters. Since the MJCI is linear in $f(x)$, it suffices to compute the
identity just for $f(x)$ of that form. In fact, using a few simple
combinatorial identities, there is a considerable
simplification of the result if we restrict $f(x)$ to be a rational function
of the variable $t$, say $t^r (1-t)^s$.
This seems a natural restriction because the correlation
functions made from intertwining operators were defined from series in $z_1$
and $z_2$ which could be written in terms of $t$. Only later did we have to
express them in terms of the local variables $x_\alpha$ in order to use
properties of the hypergeometric functions to relate the functions at the six
poles. It makes the local variables seem like a necessary but artificial
technical construction, and one might think that the final answer should
reflect the fact that there were only three possible
poles in the $t$-plane. This may
not be the last word on that subject, but our results seem to show that
with that restriction on $f$ there are six terms in the MJCI, but the second
row of the matrix yields a trivial identity. We find a slightly simpler result
if we restrict the test function to be of the form $f(x) = x^2 t^r (1-t)^s$.
We give the results in the next three corollaries.

If we were to write out each entry of
each $2\times 4$ matrix in the identity, then the result would not fit across
the page, and the pattern would be very similar in each column. So
we will just show the first column of the answer in Corollary 33.
The other columns are easily obtained from the first one by
modifying the pattern of vectors $v_1$, $v_2$, $v_3$, $v_4$ in the first row
and $\theta v_1$, $\theta v_2$, $v_3$, $v_4$ in the second row, as shown in
the matrices $[\tPhi]_\alpha$ and $[\tPsi]_\alpha$ defined before Theorem 18.
In Corollaries 34 and 35, since the second row of the matrix yields a trivial
identity, we only give the identity coming from the first row.

\pr{Corollary 33} Let $v_1,v_2,v_3\in \bV_1$ and $v_4\in\bV_3$ with
$wt(v_i) = |v_i| = N_i + \Delta_{n_i}$ and $N = N_1 + N_2 + N_3 - N_4$. Let
$$f(x) = x^{m_0} (1+\bi x)^{m_\bi} (1-\bi x)^{m_{-\bi}} (1-x)^{m_1}
(1+x)^{m_{-1}}$$
for $m_0, m_\bi, m_{-\bi}, m_1, m_{-1} \in {\Bbb Z}$, and let
$m = m_0 + m_\bi + m_{-\bi} + m_1 + m_{-1}$.
Define the following constants in terms of those parameters:
$$A_\infty=(-1)^{m_1+m_{-i}+1} (i)^{m_i+m_{-i}},$$
$$A_i=2^{m_1+m_i+m_{-i}+1} (i)^{m_0+m_{i}} (1+i)^{m_{-1}-m_1},$$
$$A_{-i}=2^{m_{-1}+m_i+m_{-i}+1} (-1)^{m_0+m_{-i}+1} (i)^{m_0+m_{-i}}
(1+i)^{m_{1}-m_{-1}},$$
$$A_{1}=2^{m_1+m_{-1}+m_{-i}+1} (-1)^{m_1}
(1+i)^{m_{i}-m_{-i}},$$
$$A_{-1}=2^{m_1+m_{-1}+m_i+1} (-1)^{m_0}
(1+i)^{m_{-i}-m_{i}},$$
and for each $k\geq 0$ define
$$Q_k = 2(k + N_1 - N_4), \qquad R_k = 2(k - N_1 - N_3), \qquad
S_k = 2(k - N_1 - N_2).$$
In the following formula, for each $k\geq 0$, and for each
$\alpha\in\{0,\infty,\bi,-\bi,1,-1\}$, $\sum_\alpha$ and $\sum'_\alpha$
denote summations over all integers
$p_1, p_{-1}, p_\bi, p_{-\bi} \geq 0$ such that the sum
$p_1 + p_{-1} + p_\bi + p_{-\bi}$ equals a fixed value, $\hp$, depending on $k$
and some of the given parameters. The fixed values are as follows.
For $\sum_0$, $\hp = -m_0-2-Q_k$,
for $\sum'_0$, $\hp = -m_0-1-Q_k$,
for $\sum_\infty$, $\hp = m+1-Q_k$,
for $\sum'_\infty$, $\hp = m+2-Q_k$,
for $\sum_\bi$, $\hp = -m_\bi-2-R_k$,
for $\sum'_\bi$, $\hp = -m_\bi-1-R_k$,
for $\sum_{-\bi}$, $\hp = -m_{-\bi}-2-R_k$,
for $\sum'_{-\bi}$, $\hp = -m_{-\bi}-1-R_k$,
for $\sum_1$, $\hp = -m_1-2-S_k$,
for $\sum'_1$, $\hp = -m_1-1-S_k$,
for $\sum_{-1}$, $\hp = -m_{-1}-2-S_k$,
and for $\sum'_{-1}$, $\hp = -m_{-1}-1-S_k$.
Within the summations we use the notations
$$C_1 = {\sfr + m_1 \choose p_1},\ C_{-1} = {\sfr +m_{-1}\choose p_{-1}},
\ C_\bi = {\sfr +m_i \choose p_i},
\ C_{-\bi} = {\sfr +m_{-i} \choose p_{-i}},$$
and
$$D = (-1)^{p_1+p_{-i}} (i)^{p_i+p_{-i}}.$$

To shorten our notation for the correlation functions, we write
$$\Phi_k = \Phi_k(v_1,v_2,v_3,v_4) =
\left(Y_{k-N_4+\svx}(v_1)Y_{-k+N_3-\svx}(v_2)v_3,v_4\right),$$
$$\Phi_k^\theta = \Phi_k(\theta v_1,\theta v_2,v_3,v_4) =
\left(Y_{k-N_4-\six}(\theta v_1)Y_{-k+N_3+\six}(\theta v_2)v_3,v_4\right),$$
$$\Phi_k^{*} = \Phi_k(v_2,v_1,v_3,v_4) =
\left(Y_{k-N_4+\svx}(v_2)Y_{-k+N_3-\svx}(v_1)v_3,v_4\right),$$
$$\Phi_k^{*\theta} = \Phi_k(\theta v_2,\theta v_1,v_3,v_4) =
\left(Y_{k-N_4-\six}(\theta v_2)Y_{-k+N_3+\six}(\theta v_1)v_3,v_4\right),$$
$$\Psi_k = \Psi_k(v_1,v_2,v_3,v_4) =
\left(Y_{N_3-N_4}(Y_{-k+N_2-\svx}(v_1)v_2)v_3,v_4\right),$$
$$\Psi_k^\theta = \Psi_k(\theta v_1,v_2,v_3,\theta v_4) =
\left(Y_{N_3-N_4}(Y_{-k+N_2+\six}(\theta v_1)v_2)v_3,\theta v_4\right).$$

Then we have the identity
$$
0 = \smK \bigg( 2^{Q_k}
\bmatrix
\sum_0
C_1 C_{-1}
{\sfr+m_{i}-Q_k-1 \choose p_i}
{\sfr+m_{-i}-Q_k-1 \choose p_{-i}}
D 2\Phi_k \\
\sum'_0
C_1 C_{-1}
{\sfr+m_{i}-Q_k \choose p_i}
{\sfr+m_{-i}-Q_k \choose p_{-i}}
D \Phi_k^\theta
\endbmatrix
$$
$$
+ 2^{Q_k} A_\infty B_\infty
\bmatrix
\sum_\infty
C_1 C_{-1}
{\sfr+m_{-i}-Q_k-1 \choose p_i}
{\sfr+m_{i}-Q_k-1 \choose p_{-i}}
D 2\Phi_k \\
\sum'_\infty
C_1 C_{-1}
{\sfr+m_{-i}-Q_k \choose p_i}
{\sfr+m_{i}-Q_k \choose p_{-i}}
D \Phi_k^\theta
\endbmatrix
$$

$$
+ 2^{R_k} A_i B_i
\bmatrix
\sum_\bi
C_1 C_{-1}
{\sfr-m-R_k-4 \choose p_i}
{\sfr+m_0-R_k-1 \choose p_{-i}}
D 2\Phi_k^{*} \\
\sum'_\bi
C_1 C_{-1}
{\sfr-m-R_k-3 \choose p_i}
{\sfr+m_0-R_k \choose p_{-i}}
D \Phi_k^{*\theta}
\endbmatrix
$$

$$
+ 2^{R_k} A_{-i} B_{-i}
\bmatrix
\sum_{-\bi}
C_1 C_{-1}
{\sfr+m_0-R_k-1 \choose p_i}
{\sfr-m-R_k-4 \choose p_{-i}}
D 2\Phi_k^{*} \\
\sum'_{-\bi}
C_1 C_{-1}
{\sfr+m_0-R_k \choose p_i}
{\sfr-m-R_k-3 \choose p_{-i}}
D \Phi_k^{*\theta}
\endbmatrix
$$

$$
+ 2^{S_k} A_{1} B_{1}
\bmatrix
\sum_1
C_\bi C_{-\bi}
{\sfr-m-S_k-4 \choose p_{1}}
{\sfr+m_0-S_k-1 \choose p_{-1}}
D \Psi_k \\
\sum'_1
C_\bi C_{-\bi}
{\sfr-m-S_k-3 \choose p_{1}}
{\sfr+m_0-S_k \choose p_{-1}}
D \Psi_k^\theta
\endbmatrix
$$

$$
+ 2^{S_k} A_{-1} B_{-1}
\bmatrix
\sum_{-1}
C_\bi C_{-\bi}
{\sfr+m_0-S_k-1 \choose p_1}
{\sfr-m-S_k-4 \choose p_{-1}}
D \Psi_k \\
\sum'_{-1}
C_\bi C_{-\bi}
{\sfr+m_0-S_k \choose p_1}
{\sfr-m-S_k-3 \choose p_{-1}}
D \Psi_k^\theta
\endbmatrix \bigg).
$$
\epr

\pr{Corollary 34} Let $v_1,v_2,v_3\in \bV_1$ and $v_4\in\bV_3$ with
$wt(v_i) = |v_i| = N_i + \Delta_{n_i}$ and $N = N_1 + N_2 + N_3 - N_4$.
For any $r,s \in {\Bbb Z}$, define
$$a = \sfr + 2r - 4,\qquad b = r + s - 1, \qquad c = -s - 1.$$
With notation as in Corollary 33, for any $r,s \in {\Bbb Z}$, we have
$$\eqalign{
0 &= 2^{2r+2(N_1-N_4)}\smK 2^{-2k} \Phi_k
\smQ {\sfr + 2s \choose  q}
{\sfr -2(r+s)-Q_k-1 \choose -Q_k/2-r-1-q} (-1)^q  \cr
&-2^{2r+2(N_1-N_4)}\smK 2^{-2k} \Phi^\theta_k
\smQ {\sfr+2s \choose q}
{\sfr - 2(r+s)-Q_k \choose -Q_k/2-r+1-q} (-1)^q \cr
&+(-1)^s 2^{-2b-2(N_1+N_3)}\smK 2^{-2k} \Phi^*_k \cdot \cr
&\qquad\qquad\smQ {\sfr+2s \choose q}
\bigg[ {a-R_k \choose b-R_k/2-q}
-3{a-R_k \choose b-R_k/2-1-q} \bigg] (-1)^q\cr
&-(-1)^s 2^{-2b-2(N_1+N_3)}\smK 2^{-2k}
\Phi^{*\theta}_k \cdot \cr
&\qquad\qquad\smQ {\sfr+2s \choose q}
\bigg[ {a-R_k+1 \choose b-R_k/2-1-q}
-3{a-R_k+1 \choose b-R_k/2-q} \bigg] (-1)^q \cr
&+2^{-2c-2(N_1+N_2)}\smK 2^{-2k} \Psi_k \cdot\cr
&\qquad\qquad\smQ {\sfr-2(r+s) \choose c-S_k/2-q}
\bigg[ {a-S_k \choose q}
-3{a-S_k \choose q-1} \bigg] (-1)^q  \cr
&-2^{-2c-2(N_1+N_2)}\smK 2^{-2k} \Psi^\theta_k \cdot\cr
&\qquad\qquad\smQ {\sfr-2(r+s) \choose c-S_k/2-q}
\bigg[ {a-S_k+1 \choose q-1}
-3{a-S_k+1 \choose q} \bigg] (-1)^q. \cr}$$
\epr

\pr{Corollary 35} Let $v_1,v_2,v_3\in \bV_1$ and $v_4\in\bV_3$ with
$wt(v_i) = |v_i| = N_i + \Delta_{n_i}$, $N = N_1 + N_2 + N_3 - N_4$ and
$\Gamma = N_1 + N_2 + N_3 + N_4$.
With notation as in Corollary 33, for any $r,s \in {\Bbb Z}$, we have

$$\eqalign{
&\smK 2^{-2k} \sum_{0 \leq q \leq k} {\sfr + 2r + 2(N_1+N_2) \choose
 q} (-1)^q \cdot \cr
&\bigg[ {\sfr - 2r+S_k-1 \choose k-q}
[ \Phi_{r+s+\Gamma-k}
+ (-1)^{r+N_1+N_2+1} \Phi^*_{-s-1-k} ] \cr
&- {\sfr - 2r+S_k \choose k-q}
[ \Phi^\theta_{r+s+\Gamma-k}
+ (-1)^{r+N_1+N_2+1}
\Phi^{*\theta}_{-s-1-k}] \bigg] \cr}$$
$$\eqalign{
&=\smK 2^{-2k} \sum_{0 \leq q \leq k} {\sfr + 2s + 2(N_1+N_3) \choose
 k-q} (-1)^q \cdot \cr
&\bigg[ {\sfr - 2s+R_k \choose q}
\Psi^\theta_{-r-1-k}
- {\sfr - 2s+R_k-1 \choose q}
\Psi_{-r-1-k}\bigg].\cr}$$
\epr
\centerline{Tables of Constants from Section 4}
$$\alignat{10}
&&n_1 &= 1&&&  n_2 &= 1&&&&&n_1 &= 3&&&  n_2 &= 3&& \\
&n_3\quad&0&\quad&1&\quad&2&\quad&3&\qquad\qquad&
&n_3\quad&0&\quad&1&\quad&2&\quad&3&  \\
&a\quad&0&\quad&\fif&\quad&0&\quad&\sfr&\qquad\qquad&
&a\quad&0&\quad&\sfr&\quad&0&\quad&\fif&  \\
&b\quad&1&\quad&\stf&\quad&\stt&\quad&\stf&\qquad\qquad&
&b\quad&1&\quad&\stf&\quad&\stt&\quad&\stf&  \\
&c\quad&0&\quad&\sth&\quad&\sot&\quad&\shf&\qquad\qquad&
&c\quad&0&\quad&\shf&\quad&\sot&\quad&\sth&  \\
&a'\quad&0&\quad&\fif&\quad&0&\quad&\sfr&\qquad\qquad&
&a'\quad&0&\quad&\sfr&\quad&0&\quad&\fif&  \\
&b'\quad&1&\quad&\stf&\quad&\stt&\quad&\stf&\qquad\qquad&
&b'\quad&1&\quad&\stf&\quad&\stt&\quad&\stf&  \\
&c'\quad&2&\quad&\sth&\quad&\sft&\quad&\sth&\qquad\qquad&
&c'\quad&2&\quad&\sth&\quad&\sft&\quad&\sth&  \\
&A\quad&-\ste&\quad&\shf&\quad&\sei&\quad&0&\qquad\qquad&
&A\quad&-\ste&\quad&0&\quad&\sei&\quad&\shf&  \\
&B\quad&0&\quad&-\ste&\quad&\shf&\quad&\sei&\qquad\qquad&
&B\quad&0&\quad&\sei&\quad&\shf&\quad&-\ste&  \\
&C\quad&-\ste&\quad&-\ste&\quad&-\ste&\quad&-\ste&\qquad\qquad&
&C\quad&-\ste&\quad&-\ste&\quad&-\ste&\quad&-\ste&  \\
&A'\quad&0&\quad&\shf&\quad&1&\quad&\shf&\qquad\qquad&
&A'\quad&0&\quad&\shf&\quad&1&\quad&\shf&  \\
&B'\quad&-\ste&\quad&-\ste&\quad&-\ste&\quad&-\ste&\qquad\qquad&
&B'\quad&-\ste&\quad&-\ste&\quad&-\ste&\quad&-\ste&  \\
&C'\quad&0&\quad&-\ste&\quad&\shf&\quad&\sei&\qquad\qquad&
&C'\quad&0&\quad&\sei&\quad&\shf&\quad&-\ste&
\endalignat$$

$$\alignat{10}
&&n_1 &= 1&&&  n_2 &= 3&&&&&n_1 &= 3&&&  n_2 &= 1&& \\
&n_3\quad&0&\quad&1&\quad&2&\quad&3&\qquad\qquad&
&n_3\quad&0&\quad&1&\quad&2&\quad&3&  \\
&a\quad&0&\quad&\sfr&\quad&-1&\quad&\sfr&\qquad\qquad&
&a\quad&0&\quad&\sfr&\quad&-1&\quad&\sfr&  \\
&b\quad&1&\quad&-\sfr&\quad&-\sot&\quad&\stf&\qquad\qquad&
&b\quad&1&\quad&\stf&\quad&-\sot&\quad&-\sfr&  \\
&c\quad&0&\quad&\shf&\quad&\sot&\quad&\sth&\qquad\qquad&
&c\quad&0&\quad&\sth&\quad&\sot&\quad&\shf&  \\
&a'\quad&0&\quad&\sfr&\quad&-1&\quad&\sfr&\qquad\qquad&
&a'\quad&0&\quad&\sfr&\quad&-1&\quad&\sfr&  \\
&b'\quad&1&\quad&\stf&\quad&-\sot&\quad&-\sfr&\qquad\qquad&
&b'\quad&1&\quad&-\sfr&\quad&-\sot&\quad&\stf&  \\
&c'\quad&2&\quad&\shf&\quad&-\stt&\quad&\shf&\qquad\qquad&
&c'\quad&2&\quad&\shf&\quad&-\stt&\quad&\shf&  \\
&A\quad&\sei&\quad&0&\quad&-\ste&\quad&\shf&\qquad\qquad&
&A\quad&\sei&\quad&\shf&\quad&-\ste&\quad&0&  \\
&B\quad&0&\quad&\sei&\quad&\shf&\quad&-\ste&\qquad\qquad&
&B\quad&0&\quad&-\ste&\quad&\shf&\quad&\sei&  \\
&C\quad&\sei&\quad&\sei&\quad&\sei&\quad&\sei&\qquad\qquad&
&C\quad&\sei&\quad&\sei&\quad&\sei&\quad&\sei&  \\
&A'\quad&0&\quad&0&\quad&0&\quad&0&\qquad\qquad&
&A'\quad&0&\quad&0&\quad&0&\quad&0&  \\
&B'\quad&\sei&\quad&\sei&\quad&\sei&\quad&\sei&\qquad\qquad&
&B'\quad&\sei&\quad&\sei&\quad&\sei&\quad&\sei&  \\
&C'\quad&0&\quad&-\ste&\quad&\shf&\quad&\sei&\qquad\qquad&
&C'\quad&0&\quad&\sei&\quad&\shf&\quad&-\ste&
\endalignat$$

\enddocument
\bye